\documentclass{aa}  

\usepackage[varg]{txfonts}
\usepackage{natbib}
\usepackage{graphicx}
\usepackage{dcolumn}
\usepackage{bm}
\usepackage{amssymb,amsmath,bm}
\usepackage{color}
\usepackage[colorlinks,linkcolor=red,citecolor=blue,urlcolor=blue ]{hyperref}
\usepackage{multirow}
\usepackage[utf8]{inputenc}
\usepackage{enumitem}

\makeatletter
\renewcommand*\aa@pageof{, page \thepage{} of \pageref*{LastPage}}

\newcommand{\rd}{\mathrm{d}}

\graphicspath{%
        {./figures/} %
}       

\begin{document}

\title{Information content in mean pairwise velocity and mean relative velocity between pairs in a triplet}
\titlerunning{Information content in two- and three-point mean relative velocity}

\author{Joseph Kuruvilla \and Nabila Aghanim} %\inst{1}

\offprints{Joseph Kuruvilla, \email{joseph.kuruvilla@universite-paris-saclay.fr}}

\institute{Universit\'e Paris-Saclay, CNRS,  Institut d'Astrophysique Spatiale, 91405, Orsay, France}

\date{Received Feb 2021; accepted July 2021}
\abstract{
Velocity fields provide a complementary avenue to constrain cosmological information, either through the peculiar velocity surveys or the kinetic Sunyaev Zel'dovich effect. One of the commonly used statistics is the mean radial pairwise velocity. Here, we consider the three-point mean relative velocity (i.e. the mean relative velocities between pairs in a triplet). Using halo catalogs from the Quijote suite of \textit{N}-body simulations, we first showcase how the analytical prediction for the mean relative velocities between pairs in a triplet achieve better than 4-5\% accuracy using standard perturbation theory at leading order for triangular configurations with a minimum separation of $r \geq 50\ h^{-1}\mathrm{Mpc}$. Furthermore, we present the mean relative velocity between pairs in a triplet as a novel probe of neutrino mass estimation. We explored the full cosmological information content of the halo mean pairwise velocities and the mean relative velocities between halo pairs in a triplet. We did this through the Fisher-matrix formalism using 22,000 simulations from the Quijote suite and by considering all triangular configurations with a minimum and a maximum separation of $20\ h^{-1}\mathrm{Mpc}$ and $120\ h^{-1}\mathrm{Mpc}$, respectively. We find that the mean relative velocities in a triplet allows a 1$\sigma$ neutrino mass ($M_\nu$) constraint of 0.065 eV, which is roughly 13 times better than the mean pairwise velocity constraint (0.877 eV). This information gain is not limited to neutrino mass, but it extends to other cosmological parameters: $\Omega_{\mathrm{m}}$, $\Omega_{\mathrm{b}}$, $h$, $n_{\mathrm{s}}$, and $\sigma_{8}$, achieving an information gain of 8.9, 11.8, 15.5, 20.9, and 10.9 times, respectively. These results illustrate the possibility of exploiting the mean three-point relative velocities to constrain the cosmological parameters accurately from future cosmic microwave background experiments and peculiar velocity surveys.}


\defcitealias{Planck2018}{Planck~Collaboration~2020}
\defcitealias{Planck16}{Planck~Collaboration~2016}
\defcitealias{Planck13:cosmology}{Planck~Collaboration~2014}
\defcitealias{eBOSS20}{eBOSS~Collaboration~2020}

\maketitle
%%%%%%%%%%%%%%%%%%%%%%%%
\section{Introduction}
%%%%%%%%%%%%%%%%%%%%%%%%

The last decades have witnessed a tremendous increase in our understanding of the underlying cosmological model. This has been mainly facilitated by cosmic microwave background (CMB) experiments \citepalias[e.g.][]{Planck2018}, galaxy redshift surveys \citepalias[e.g.][]{eBOSS20}, and gravitational lensing surveys \citep[e.g.][]{Kids1000}. However, there are several key areas in which our understanding is incomplete or lacking. The questions regarding the nature of dark matter and dark energy and the summed neutrino mass ($M_\nu$) are the two main scientific drivers for forthcoming redshift surveys such as  Euclid\footnote{\url{https://www.euclid-ec.org/}} and future CMB experiments such as the Simons Observatory\footnote{\url{https://simonsobservatory.org/}} \cite[SO,][]{SO} and CMB-S4\footnote{\url{https://cmb-s4.org/}} \citep{CMBS4}.

Cosmological information from galaxy redshift surveys is often extracted using the two-point correlation function in the configuration space or its Fourier analogue, the power spectrum. However, as the complex galaxy distribution represents a non-Gaussian field, higher order statistics such as the bispectrum and the three-point correlation function contain additional information. Thus, by considering both two-point and three-point clustering information, future redshift surveys will be able to obtain a substantial improvement with regard to the cosmological parameter constraints \citep[e.g.][]{YankelevichPorciani18, ChudaykinIvanov19, Gualdi+20, Agarwal+20, Samushia+21}. The determination of the summed mass of neutrinos will be one of the primary goals of future redshift surveys. This will be possible as the free-streaming of neutrinos imprint unique signatures on galaxy clustering information in both real and redshift space \citep[e.g.][]{Saito+08,Wong08,Castorina+15,Navarro+18,Garcia+19,KamalinejadSlepian20}.

Complementary to clustering statistics based on the density field is the study of summary statistics based on the peculiar velocity field.  We can observe the velocity fields via: (i) peculiar velocity surveys, and (ii) the kinetic Sunyaev-Zel'dovich (kSZ) effect \citep{SZ72, SZ80}.  Peculiar velocity can be directly measured using empirical relations like the Tully-Fisher \citep{TullyFisher77} relation and the fundamental plane \citep{DjorgovskiDavis87, Dressler+87} relation. These surveys offer an opportunity to probe the peculiar velocities in the nearby Universe. In particular, future peculiar velocity surveys like the Taipan galaxy survey\footnote{\url{https://www.taipan-survey.org}} \citep{Taipan}, and the Widefield ASKAP  L-band  Legacy  All-sky  Blind Survey\footnote{\url{https://www.atnf.csiro.au/research/WALLABY/}} \cite[WALLABY,][]{Wallaby-20} have been shown to constrain the growth rate of structure rather competitively with the current galaxy redshift surveys at low redshift \citep{Koda+14, Howlett+17}. Recently, \cite{Dupuy+19} measured the growth rate of the structure by applying the mean pairwise velocity estimator on the Cosmicflows-3 dataset \citep{Cosmicflows3}.

The kSZ effect is a secondary anisotropy where CMB photons are scattered off free electrons in motion due to the bulk motion of the ionised medium; e.g. cluster peculiar motion. It results in a Doppler shift, which preserves the blackbody spectrum of the CMB. The first significant detection of the kSZ effect was achieved through the mean pairwise velocity statistics \citep{hand+12} using an estimator developed in \cite{Ferreira+99}. Further detection of the kSZ effect using the mean pairwise velocities were presented in \citealt{Carlos+15},~\citetalias{Planck16},~\citealt{Soergel+16},~\citealt{Bernardis+17},~\citealt{Li+17}, and \citealt{Calafut+21}. The mean radial pairwise velocity has been shown to be capable of constraining alternative theories of gravity and dark energy \citep{BhattacharyaKosowsky07,BhattacharyaKosowsky08,KosowskyBhattacharya09,Mueller+15a}.  There have been other methods via which the kSZ effect has been detected;  for example, (i) by correlating CMB maps with reconstructed velocity field \cite[e.g.][]{Carlos+15,Schaan+16,Nguygen+20,Tanimura+20}, and (ii) by cross correlating CMB maps with angular redshift fluctuation maps \citep{Chaves-Montero+19}. The forthcoming CMB surveys including SO, CMB-S4, and CMB-HD \citep{CMBHD-I, CMBHD-II} will offer a plethora of data and allow us to measure kSZ effect accurately.

In addition to their imprint on galaxy clustering, massive neutrinos also leave their imprint on various velocity statistics. For the pairwise velocity statistics, the impact of massive neutrinos and its interplay with baryonic physics at non-linear scales was studied in \cite{Kuruvilla+20} using the BAHAMAS suite of simulations \citep{McCarthy+18}. It was found that the mean matter pairwise velocity decreases with increasing neutrino mass at those pair separations. For analytical predictions, \cite{AvilesBanerjee20} studied the effect of massive neutrinos on pairwise velocity using Lagrangian perturbation theory for biased tracers at quasi-linear scales and above. The pairwise velocity measurement from kSZ experiments has been shown to be a novel probe to constrain the summed neutrino mass \citep{Mueller+15b}, and provide complementary constraints with respect to galaxy clustering and CMB experiments. In line with this, we want to explore how much information the higher-order relative velocity statistics provide on the cosmological parameters, in particular with regard to summed neutrino mass. The three-point generalisation of the pairwise velocity, i.e. the relative velocities between pairs in a triplet, was recently introduced in \cite{KuruvillaPorciani20}. However, the impact of cosmological parameters (especially massive neutrinos) on it remains unexplored. Thus, in this paper, we study how the different cosmological parameters (including $M_\nu$) affect the relative velocities between pairs in a triplet and this constitutes one of the main aims of this work.

We used the Fisher-matrix formalism to quantify the viability of using the mean relative velocities between pairs in a triplet to constrain cosmological parameters, and we compared it with the cosmological information obtained from the mean pairwise velocity statistics. For this purpose, we measured the necessary derivatives and the covariance matrix directly from the Quijote suite of simulations \citep{Quijote20}.  Thus, this work forms a parallel line of study to \cite{Hahn+20} and \cite{HahnVilla20}, which quantified the information content in the redshift-space bispectrum of halos and galaxies, respectively, using the Fisher-matrix formalism; moreover, they computed the necessary quantities directly from the Quijote simulations. It was shown that the redshift-space bispectrum was able to achieve sizeable gains in the cosmological parameter constraints over those obtained from the power spectrum; notably, it was especially able to constrain summed neutrino mass to high precision. We aim to study if the mean relative velocity statistics can provide competitive constraints when compared to those from the clustering analyses.

The paper is structured as follows. The Quijote suite of simulation, which we used in this work, is introduced briefly in Sect.~\ref{sec:sims}. The information content of the velocity statistics was studied using the Fisher-matrix formalism, which is briefly summarised in Sect.~\ref{sec:fisher}. In Sect.~\ref{sec:velsta}, we describe the summary statistics based on velocity that we employed in this work. Our results are presented in Sect.~\ref{sec:results} and finally concluded in Sect.~\ref{sec:conclusions}.

\section{Data and analysis}
%%%%%%%%%%%%%%%%%%%%%%%%
\subsection{Quijote simulation suite}
\label{sec:sims}
%%%%%%%%%%%%%%%%%%%%%%%%

The Quijote\footnote{\url{https://quijote-simulations.readthedocs.io/}} \citep{Quijote20} simulation suite contains 44,100 \textit{N}-body simulations spanning more than a few thousand cosmological models, run using the tree-PM code \textsc{gadget}-3 \citep{Springel05}. The initial conditions (ICs) of the simulation, at redshift $z=127$, were generated using the second-order Lagrangian perturbation theory (for runs with zero mass neutrino cosmology) and the Zel'dovich approximation (for simulations with massive neutrinos). Each of these simulations spans a volume of $1\,h^{-3}\mathrm{Gpc}^3$ and follows the cosmological evolution of $512^3$ cold dark matter (CDM) particles (and, additionally, in the  case of massive neutrino simulations, $512^3$ neutrino particles). The fiducial cosmological parameters for the simulation are consistent with the \citetalias{Planck2018}, and these are as follows: (total matter density) $\Omega_{\mathrm{m}}=0.3175$, (baryonic matter density) $\Omega_{\mathrm{b}}=0.049$, (primordial spectral index of the density perturbations) $n_{\mathrm{s}}=0.9624$, (amplitude of the linear power spectrum on the scale of $8\ h^{-1}\mathrm{Mpc}$) $\sigma_8=0.834$,  and (present-day value of the Hubble constant) $H_0\equiv H(z=0)=100\, h\,\mathrm{km}\,\mathrm{s}^{-1}\mathrm{Mpc}^{-1}$ with $h=0.6711$. However, it assumes neutrinos to be massless, i.e. $M_\nu=0.0$ eV. The reference run with the fiducial cosmological model consists of 15,000 random realisations. The Quijote suite of simulations is thus ideal for constructing an accurate covariance matrix of any cosmological observable.  Similarly, it provides a set of 500 random realisations where only one cosmological parameter is varied from the fiducial cosmology.  The variation in the cosmological parameters includes an increment and decrement in the cosmological parameter in consideration. The step size for this change utilised in the Quijote suite of simulations are as follows: $\mathrm{d}\Omega_{\mathrm{m}} =  0.010, \mathrm{d}\Omega_{\mathrm{b}} =  0.002, \mathrm{d}h =  0.020, \mathrm{d}n_{\mathrm{s}} =  0.020$, and $\mathrm{d}\sigma_8 =  0.015$. Furthermore, the suite provides 500 realisations for the three massive neutrino cosmologies, i.e. $M_\nu = 0.1, 0.2$ and $0.4$ eV. The Quijote suite provides an additional 500 random realisations for the reference cosmology, in which the ICs were generated using the Zel'dovich approximation. This allowed us to compute the numerical derivatives with respect to massive neutrinos.

In this work, we used halo catalogue data from 22,000 \textit{N}-body simulations of the Quijote suite. These halos  were identified using a friends-of-friends algorithm. We selected halos that have $M_\mathrm{h} > 5 \times 10^{13}\ h^{-1}\mathrm{M}_\odot$ (corresponding to groups and clusters of galaxies) at $z=0,$ which gives a mean number density of $\bar{n} \sim 0.92 \times 10^{-4}\,h^3\,\mathrm{Mpc}^{-3}$ for the reference simulations.  For further details regarding the Quijote suite of simulations, we refer the reader to \cite{Quijote20}.

%%%%%%%%%%%%%%%%%%%%%%%%
\subsection{Fisher-matrix formalism}
\label{sec:fisher}
%%%%%%%%%%%%%%%%%%%%%%%%
In this work, we used the Fisher-matrix formalism to quantify the error estimates on the cosmological parameters. The Fisher information matrix can be defined as \citep[e.g.][]{Tegmark+1997, Heavens09, Verde10}
\begin{equation}
    F_{\alpha \beta} = \left\langle -\frac{\partial^2\ln{\mathcal{L}}}{\partial \theta_\alpha \partial \theta_\beta} \right\rangle \, ,
    \label{eq:fisher_definition}
\end{equation}
where $\mathcal{L}$ is the likelihood of the data given a model, and $\theta_\alpha$ and $\theta_\beta$ are two of the model parameters. Under the assumption that the probability distribution of the data is a Gaussian (and hence has a Gaussian likelihood), we can write the Fisher information matrix as
\begin{equation}
    F_{\alpha \beta} = \frac{\partial\boldsymbol{S}}{\partial \theta_\alpha} \cdot \hat{\mathbf{C}}^{-1} \cdot \frac{\partial\boldsymbol{S}^\mathsf{T}}{\partial \theta_\beta} \, ,
    \label{eq:fisher_reduced}
\end{equation}
where $\boldsymbol{S}$ represents the data vector for the summary statistics we are interested in. In our case, it is the mean pairwise velocity and the mean relative velocities between pairs in a triplet. The precision matrix (i.e. the inverse covariance matrix) is given by $\hat{\mathbf{C}}^{-1}$, and to obtain it, we computed the covariance matrix of the desired statistics directly from the simulations as follows:
\begin{equation}
    \widetilde{\mathbf{C}} = \frac{1}{N_{\mathrm{sims}}-1}\sum_{i=1}^{N_{\mathrm{sims}}} \left(\boldsymbol{S}_i-\overline{\boldsymbol{S}}\right)\left(\boldsymbol{S}_i-\overline{\boldsymbol{S}}\right)^\mathsf{T} \, ,
    \label{eq:covariancematrix}
\end{equation}
where $\overline{\boldsymbol{S}} = \frac{1}{N_{\mathrm{sims}}}\sum_{i=1}^{N_{\mathrm{sims}}} \boldsymbol{S}_i$. The total number of simulations used to compute the covariance matrix is denoted by $N_{\mathrm{sims}}$, and in our case $N_{\mathrm{sims}}=15,000$. An unbiased estimate of the covariance matrix is obtained using Eq.~\ref{eq:covariancematrix}. However, its inversion leads to a biased estimate of the precision matrix. It can be corrected statistically by a multiplicative correction factor \citep{Kaufmann67,Anderson03,Hartlap+07}:
\begin{equation}
    \hat{\mathbf{C}}^{-1} = \frac{N_{\mathrm{sims}}-N_{\mathrm{bins}}-2}{N_{\mathrm{sims}}-1}\ \widetilde{\mathbf{C}}^{-1} \, ,
\end{equation}
where $N_{\mathrm{bins}}$ is the number of bins in the data vector. It should be noted that by using Eq.~\ref{eq:fisher_reduced}, we assumed that the covariance matrix is independent of cosmology. 
However, the correction due to the cosmology dependence of the covariance matrix is expected to be negligible \citep{Kodwani+19}.

The derivatives required to construct the Fisher information matrix were numerically computed using the Quijote suite of simulations. In the case when the model parameter $\theta \equiv \{\Omega_{\mathrm{m}}, \Omega_{\mathrm{b}}, h, n_{\mathrm{s}}, \sigma_{8}\}$, we made use of the central difference approximation:
\begin{equation}
    \frac{\partial \boldsymbol{S}}{\partial \theta} \simeq \frac{\boldsymbol{S}(\theta+\mathrm{d}\theta)-\boldsymbol{S}(\theta-\mathrm{d}\theta)}{2\ \mathrm{d}\theta} \, .
\end{equation}
As mentioned in Sect. \ref{sec:sims}, for each cosmological parameter considered the Quijote suite provides 500 realisations where only one parameter is varied, while the rest are fixed at the fiducial cosmological model. This enables the derivatives to be calculated numerically.

In the case of the neutrino mass, the fiducial value is 0.0 eV, and it cannot have negative values; hence, we obtained the partial derivative using
\begin{equation}
    \frac{\partial \boldsymbol{S}}{\partial M_\nu} \simeq \frac{-\boldsymbol{S}(M_\nu=0.4)+4\boldsymbol{S}(M_\nu=0.2) - \boldsymbol{S}(M_\nu=0)}{0.4} \, .
\end{equation}
Thus, we utilised two sets of massive neutrino simulations from Quijote, with $M_\nu = 0.2$ eV and $M_\nu = 0.4$ eV for the Fisher information matrix. These simulations had the initial conditions generated using Zel'dovich approximation. For the case of $M_\nu = 0$ eV for the above derivative, to be consistent, we hence made use of another 500 realisations of reference cosmology simulation  in which the initial conditions were also similarly generated using the Zel'dovich approximation.

%%%%%%%%%%%%%%%%%%%%%%%%
\section{Mean relative velocity statistics}
\label{sec:velsta}
%%%%%%%%%%%%%%%%%%%%%%%%

In the single stream fluid approximation, the mean matter pairwise velocity can be defined as 
\begin{align}
\langle \bm{w}_{12}|\bm{r}_{12} \rangle_\mathrm{p}   = &\ \displaystyle \frac{\langle(1+\delta_{1})(1+\delta_{2}) (\bm{v}_{2}-\bm{v}_{1})\rangle}{\langle (1+\delta_{1})(1+\delta_{2})\rangle} \, ,
 \label{eq:mean-radial-velocity-two}
 \end{align}
where $\delta_i\equiv\delta(\bm{x}_i)$ is the density contrast, and $\bm{v}_i\equiv \bm{v}(\bm{x}_i)\equiv \bm{u}(\bm{x}_i)/aH$ is the normalised peculiar velocity (where $a$ is the scale factor). At leading order in standard perturbation theory, we have $\langle \bm{w}_{12}|\bm{r}_{12} \rangle_\mathrm{p} 
 \simeq \langle \delta_{1}\bm{v}_{2} \rangle - \langle \delta_{2} \bm{v}_{1} \rangle$. Thus, the mean pairwise velocity can be written as \citep[e.g.][]{Fisher95,juszkiewicz+98, ReidWhite11}
\begin{equation}
\langle \bm{w}_{12}|\bm{r}_{12} \rangle_{\mathrm p}  
 \simeq \displaystyle - \frac{f}{\pi^2} \, \hat{\bm{r}}_{12} \int_0^{\infty} k \,j_1(k r_{12})\,P(k)\, \rd k = \bar{w}(r_{12})\,\hat{\bm{r}}_{12}\, ,
 \label{eq:mean-radial-velocity}
\end{equation}
where $j_1(x) = \sin (x)/x^2- \cos (x)/x$, $P(k)$ denotes the linear matter power spectrum, the subscript p implies that the averages are computed over all particle pairs with separation $r_{12}$, and the symbol $\simeq$ indicates that the mathematical expression was truncated to leading order. It should be noted that due to gravity, on average the particles in a pair approach each other, i.e., $\bar{w}(r_{12})<0$. The fidelity of this analytical prediction was tested against simulations (for both dark matter particles and halos), and we found that it  reproduces the measurements well above pair separations of 40-50 $h^{-1}\mathrm{Mpc}$ \citep[e.g.][]{ReidWhite11,KuruvillaPorciani18}. Improvements have been made on the leading order prediction from standard perturbation theory by incorporating different flavours of perturbation theory, such as convolutional Lagrangian perturbation theory \citep[CLPT,][]{Carlson+13, Wang+14} and  convolutional Lagrangian effective field theory \citep[CLEFT,][]{Vlah+16}. Accurate perturbation theory prediction accounting for massive neutrinos was introduced in \cite{AvilesBanerjee20} and in the case of modified gravity in \cite{ValogiannisBean20}. In parallel, advances in semi-analytical approaches to modelling mean pairwise velocity has been shown to achieve 10-20\% precision at low redshift for fairly non-linear scales of $5\ h^{-1}\mathrm{Mpc}$ and above \citep{Shirasaki+20}.

The relative velocity between pairs can be generalised to triplets with separations $\triangle_{123}=(r_{12},r_{23},r_{31})$ and is valid for all tracers (e.g. dark matter particles, halos, and galaxies). This was first done for dark matter in \cite{KuruvillaPorciani20}. In such a case, we consider two mean relative velocities:
$\langle \bm{w}_{12}|\triangle_{123} \rangle_\mathrm{t}$ and
$\langle \bm{w}_{23}|\triangle_{123} \rangle_\mathrm{t}$
(where the subscript t implies that the averages are computed over all particle triplets with separations $\triangle_{123}$). The relative velocity between pair 12 in a triplet, in the single stream fluid approximation, can be written as
\begin{align}
\langle \bm{w}_{12}|\triangle_{123} \rangle_\mathrm{t}   = &\ \displaystyle \frac{\langle(1+\delta_{1})(1+\delta_{2}) (1+\delta_{3})(\bm{v}_{2}-\bm{v}_{1})\rangle}{\langle (1+\delta_{1})(1+\delta_{2}) (1+\delta_{3})\rangle} \nonumber \\
 \simeq &\ \langle \delta_{1}\bm{v}_{2} \rangle - \langle \delta_{2} \bm{v}_{1} \rangle + \langle \delta_{3}\bm{v}_{2} \rangle - \langle \delta_{3} \bm{v}_{1} \rangle\nonumber \\
 =&\ \bar{w}(r_{12})\,\hat{\bm{r}}_{12}-\frac{1}{2}\left[
 \bar{w}(r_{23})\,\hat{\bm{r}}_{23}+\bar{w}(r_{31})\,\hat{\bm{r}}_{31}\right]
 \, .
 \label{eq:mean-radial-velocity-three}
 \end{align}

Contrary to the case of mean pairwise velocities, the mean relative velocity between a particle pair in a triplet is not purely radial, it also has a transverse component in the plane of the triangle defined by the particles. The transverse component is generated by the gravitational influence of the third particle on the pair. We can decompose  the  mean relative velocity into its radial and transverse components:
\begin{align}
   \langle \bm{w}_{12}|\triangle_{123} \rangle_\mathrm{t}  & = \langle \bm{w}_{12}\cdot \hat{\bm{r}}_{12} |\triangle_{123} \rangle_{\mathrm t}\,\hat{\bm{r}}_{12} +
   \langle \bm{w}_{12}\cdot \hat{\bm{t}}|\triangle_{123} \rangle_{\mathrm t}\, \hat{\bm{t}}\nonumber\\
&   =R_{12}(\triangle_{123})\,\hat{\bm{r}}_{12}+T_{12}(\triangle_{123})\,\hat{\bm{t}}\;,
   \label{eq:decomposition}
\end{align}
where $\hat{\bm{t}}=(\hat{\bm{r}}_{23}-\cos\chi\,\hat{\bm{r}}_{12})/\sin\chi$ and $\chi= \arccos(\hat{\bm r}_{12} \cdot \hat{\bm r}_{23})$. Thus, we obtain the radial (matter) component as 
 \begin{align}
 R_{12}(\triangle_{123})=
 \bar{w}(r_{12})&-\frac{1}{2}\Bigg[
 \bar{w}(r_{23})\,\cos \chi \nonumber \\
 & -\bar{w}(r_{31})\,\frac{r_{12}+r_{23}\cos\chi}{\sqrt{r_{12}^2+r_{23}^2+2r_{12}r_{23}\cos\chi}}\Bigg]\;.
 \label{eq:R12_triangle}
 \end{align}

 \noindent Similarly, the mean radial relative velocity can be written for the pair 23 in $\triangle_{123}$:
\begin{align}
 R_{23}(\triangle_{123})=
 \bar{w}(r_{23})&-\frac{1}{2}\Bigg[
 \bar{w}(r_{12})\,\cos \chi \nonumber \\
 & -\bar{w}(r_{31})\,\frac{r_{23}+r_{12}\cos\chi}{\sqrt{r_{12}^2+r_{23}^2+2r_{12}r_{23}\cos\chi}}\Bigg]\;.
 \label{eq:R23_triangle}
 \end{align}
 
\begin{figure}
  \centering
  \includegraphics[scale=0.58]{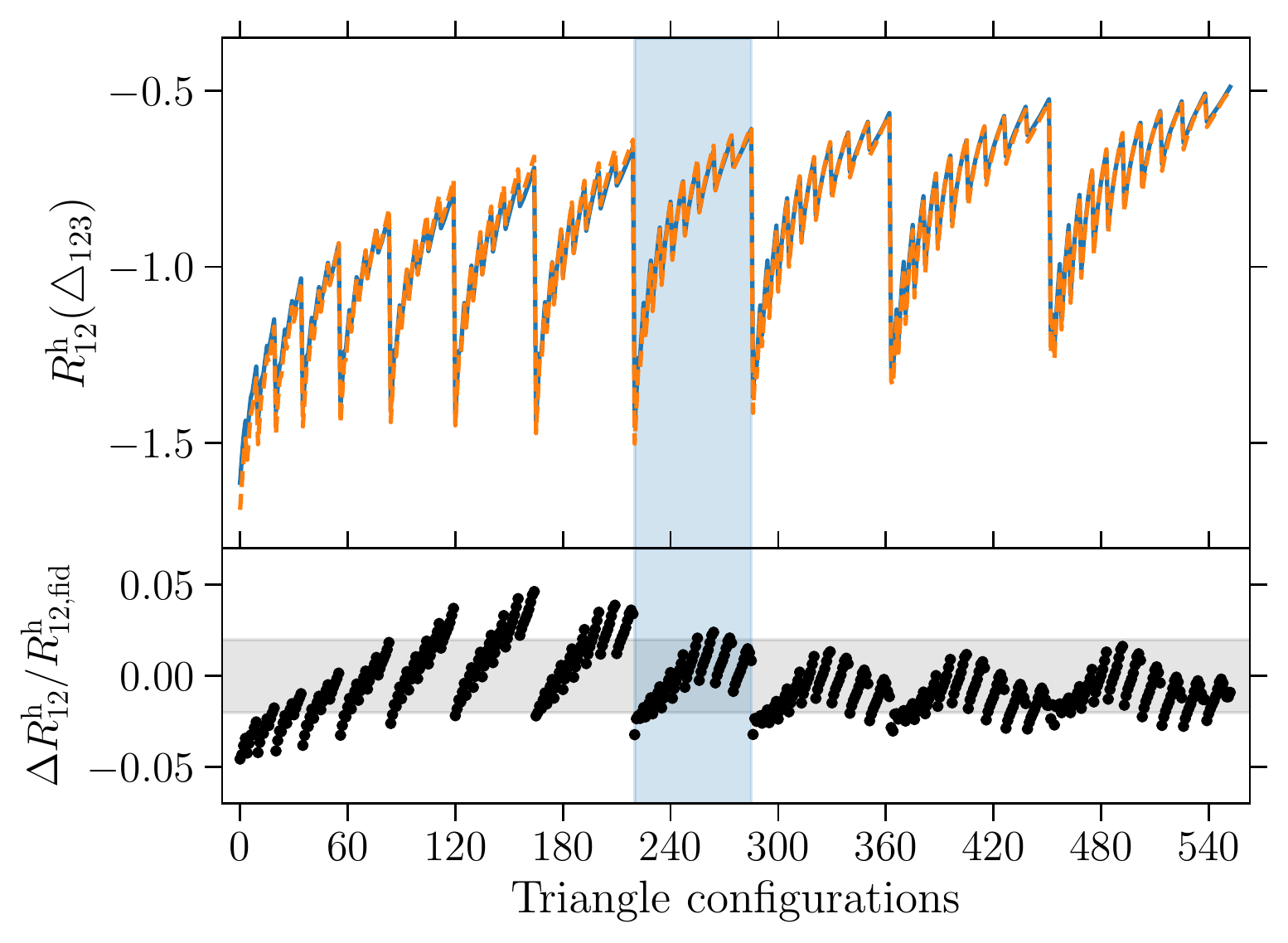} \\
  \includegraphics[scale=0.58]{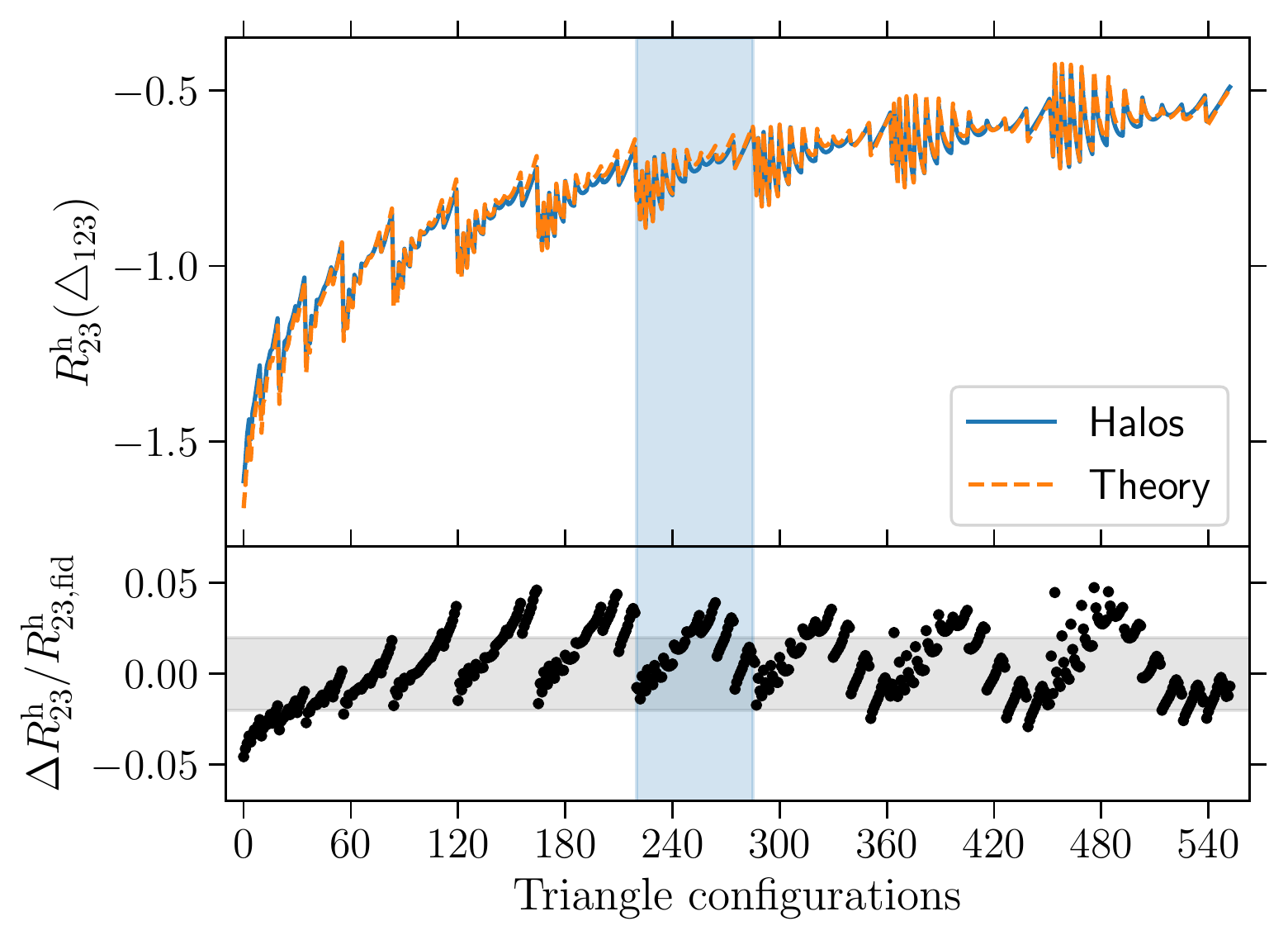}
  \caption{Top panel shows the $R^{\mathrm{h}}_{12}$ measurement (solid line) from the halo catalogues and the analytical prediction (dashed line) using linear perturbation theory as given in Eq.~\ref{eq:bias}.  Bottom panel shows the same for $R^{\mathrm{h}}_{23}$. The plot shows all triangular configurations, in which the minimum and maximum separations for a triangular side are 50 and 120 $h^{-1}\mathrm{Mpc}$, respectively. Light blue vertical band corresponds to triangular configurations where one leg of the triangle is fixed, $ r_{12} \in (100, 105)\ h^{-1}\mathrm{Mpc}$.} 
  \label{fig:mean_three_vel_halos}
\end{figure}

For detailed derivations of these mean relative velocities between pairs in a triplet, we refer the reader to  \cite{KuruvillaPorciani20}, in which we also showcase how well the standard perturbation theory prescription at leading order works for an unbiased tracer (i.e. for dark matter species). Utilising a massive number of simulations, we quantified the information content of mean relative velocity between halo pairs in a triplet.
To compare the measurement against the theoretical prediction, we employed a linear (scale-independent and mass-dependent) bias $b(\overline{M}_\mathrm{h})$, and it can be written as
\begin{align}
    R^{\mathrm{h}}_{12}(\triangle_{123}, M_\mathrm{h}) & \simeq \langle(1+b\delta_{1})(1+b\delta_{2}) (1+b\delta_{3})(\bm{v}_{2}-\bm{v}_{1})\rangle \nonumber \\ 
    & = b(\overline{M}_\mathrm{h})\,R_{12}(\triangle_{123}), 
    \label{eq:bias}
\end{align} 
and similarly
\begin{equation}
    R^{\mathrm{h}}_{23}(\triangle_{123}, M_\mathrm{h}) = b(\overline{M}_\mathrm{h})\,R_{23}(\triangle_{123}) \,,
    \label{eq:bias_23}
\end{equation}
where the superscript `h' represents the mean statistics for the halos, and in our case $\overline{M}_\mathrm{h}$ implies all halos with mass greater than $5 \times 10^{13}\ h^{-1}M_{\odot}$. To quantify the information content using the Fisher-matrix formalism, we measured the three-point mean relative velocity statistics for all triangular configurations with separation lengths $r_{12} \geq r_{23} \geq r_{31} \geq 20\ h^{-1}\mathrm{Mpc}$ and separation lengths less than $120\ h^{-1}\mathrm{Mpc}$. All separations use a bin width of 5 $h^{-1}\mathrm{Mpc}$ and thus have 1168 total triangular configurations.  

We directly computed $R^{\mathrm{h}}_{ij}$ from the simulations by measuring the average of the radial components of the relative velocities, i.e. $(\bm{v}_j - \bm{v}_i)\cdot \hat{\bm{r}}_{ij}$, and binning them to the corresponding triangles. For each halo catalogue of the Quijote simulation, this corresponds to computing the relative velocities for about a total number of $2$ billion triangles at the scales we considered. In Fig.~\ref{fig:mean_three_vel_halos}, we show the theoretical prediction alongside the direct measurements from the simulation of these relative velocities between halo pairs in a triplet. However, to plot the comparison with the theoretical prediction, we show the direct measurements of $R^\mathrm{h}_{12}$ and $R^\mathrm{h}_{23}$ for triangle configurations with $r_{12} \geq r_{23} \geq r_{31} \geq 50\ h^{-1}\mathrm{Mpc}$ and separation lengths of less than 120 $h^{-1}\mathrm{Mpc}$. We restricted the plotting to these triangular configurations for two reasons: (i) we expected the analytical prediction to work well at these quasi-linear scales and above, and (ii) we reduced the number of total triangular configurations (from 1168 to 552 configurations) considered in the plot to highlight the configurations that are most interesting for comparison purposes. As stated before for all separations, we used 5 $h^{-1}\mathrm{Mpc}$ -wide bins. Thus, the smallest triangle (triangle configuration `0') considered in Fig.~\ref{fig:mean_three_vel_halos} corresponds to ${r_{12}, r_{23}, r_{31}} \in \{(50,55), (50,55), (50,55)\} \ h^{-1}\mathrm{Mpc}$, while the largest-scale triangle (triangle configuration `552') has ${r_{12}, r_{23}, r_{31}} \in \{(115,120), (115,120), (115,120)\} \ h^{-1}\mathrm{Mpc}$.  In the top and bottom rows, we plot the relative velocity between pairs 12 and 23 in $\triangle_{123,}$ respectively. The bias factor as given in Eq.~\ref{eq:bias} was determined by fitting the prediction using least-squares method, and we obtain $b(M_\mathrm{h} > 5 \times 10^{13}) = 1.856$. The direct measurement of relative velocities from the 15,000 reference halo catalogues is denoted by the solid (blue) line, while the theoretical prediction using perturbation theory at leading order multiplied by the bias factor is given by a dashed (orange) line.  The negative values of $R_{12}$ and $R_{23}$ imply that the pairs in the triplet are infalling towards each other, on average, due to the gravitational force. The rate of infall increases as the three separation lengths of the triangle decrease. We focused on a subset of configurations with the vertical blue shaded region to highlight these effects, where we fixed one leg of the triangle ($100 < r_{12} < 105\ h^{-1}\mathrm{Mpc}$) and let $r_{23}$ and $r_{31}$ vary. It clearly shows how, contrary to the mean pairwise velocity, the addition of a third halo (or any other tracer of the matter distribution) imprints quite distinct features onto the velocity statistics. The left-most data point in the vertical shaded region has its second and third legs of the triangle fixed as $r_{23} = r_{31}\in(50,55) \ h^{-1}\mathrm{Mpc}$. Thus, as the halo `3' is closer to the pair 12, the mean velocity $R_{12}$ experiences a greater infall speed. On the other hand, in the case of $R_{23}$ for the same triangular configuration, the effect is different as the presence of halo `1' causes a reduction in the net infall velocity. The relative velocity between pairs $R_{12}$ and $R_{23}$ will be equal to each other when $r_{12} = r_{23}$, irrespective of the length of the third side. When comparing the theoretical predictions, we see that it is, overall, accurate within 4-5\%.  As can be seen from the residuals of both $R^{\mathrm{h}}_{12}$ and $R^{\mathrm{h}}_{23}$, the analytical predictions seem somewhat sensitive to the triangle shapes. This is mainly due to the fact that even for the mean halo pairwise velocity at these scales, the linear theory prediction is not perfect, and it is accurate at roughly below 2-4\% for certain scales of pair separation between 50 and 120 $h^{-1}\mathrm{Mpc}$. This has been shown in the case of mean halo pairwise velocity in Fig.~6 of \cite{Uhlemann+15}, wherein the analytical linear theory prescription over-predicts by 2-4\% at around 90-100$\ h^{-1}\mathrm{Mpc}$, and similarly under-predicts by 3-5\% at around $50\ h^{-1}\mathrm{Mpc}$. As the leading order prediction for $R^{\mathrm{h}}_{12}$ and $R^{\mathrm{h}}_{23}$ is built from the sum of different pairwise components, these small inaccuracies propagate, and we see it in the residuals. If we focus on the triangular configuration `165', we find that both $R^{\mathrm{h}}_{12}$ and $R^{\mathrm{h}}_{23}$ over-predict the measurement by 4.6\%, and this configuration corresponds to the equilateral triangle with a separation length between 90 and 95 $h^{-1}\mathrm{Mpc}$.  However, the overall degree of agreement of the analytical prediction with the direct measurement from simulation for all these configurations is quite encouraging, even with a simple biasing scheme. Hence, we believe if we were to restrict our present analysis to triangular configurations above $50\ h^{-1}\mathrm{Mpc}$, and assuming a massless neutrino cosmology, the leading order prediction would be sufficient for an analysis such as the one we undertook here. To further push to smaller non-linear scales, one would need to improve the analytical predictions for $R^{\mathrm{h}}_{ij}$ by incorporating different types of perturbation theory like CLPT or CLEFT,
as has been done in the case of mean pairwise velocities.
This provides us with the motivation to utilise direct measurements from the simulations to quantify the information content in these velocity statistics as we can extend the Fisher analysis to non-linear scales where the leading order theoretical predictions would otherwise not be accurate enough.

%%%%%%%%%%%%%%%%%%%%%%%%
\section{Results}
\label{sec:results}
%%%%%%%%%%%%%%%%%%%%%%%%
\subsection{Impact of cosmology}
\begin{figure}
  \centering
  \includegraphics[scale=0.7]{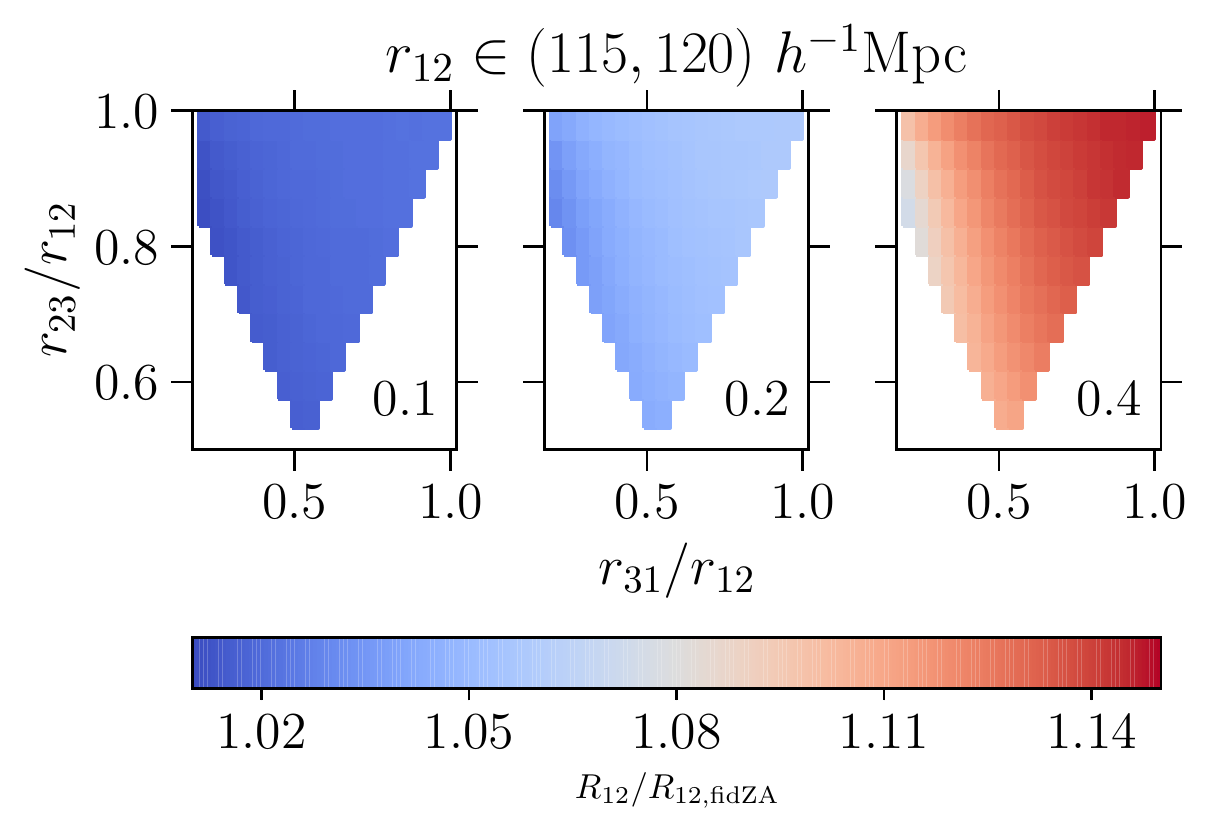} \\  
  \includegraphics[scale=0.7]{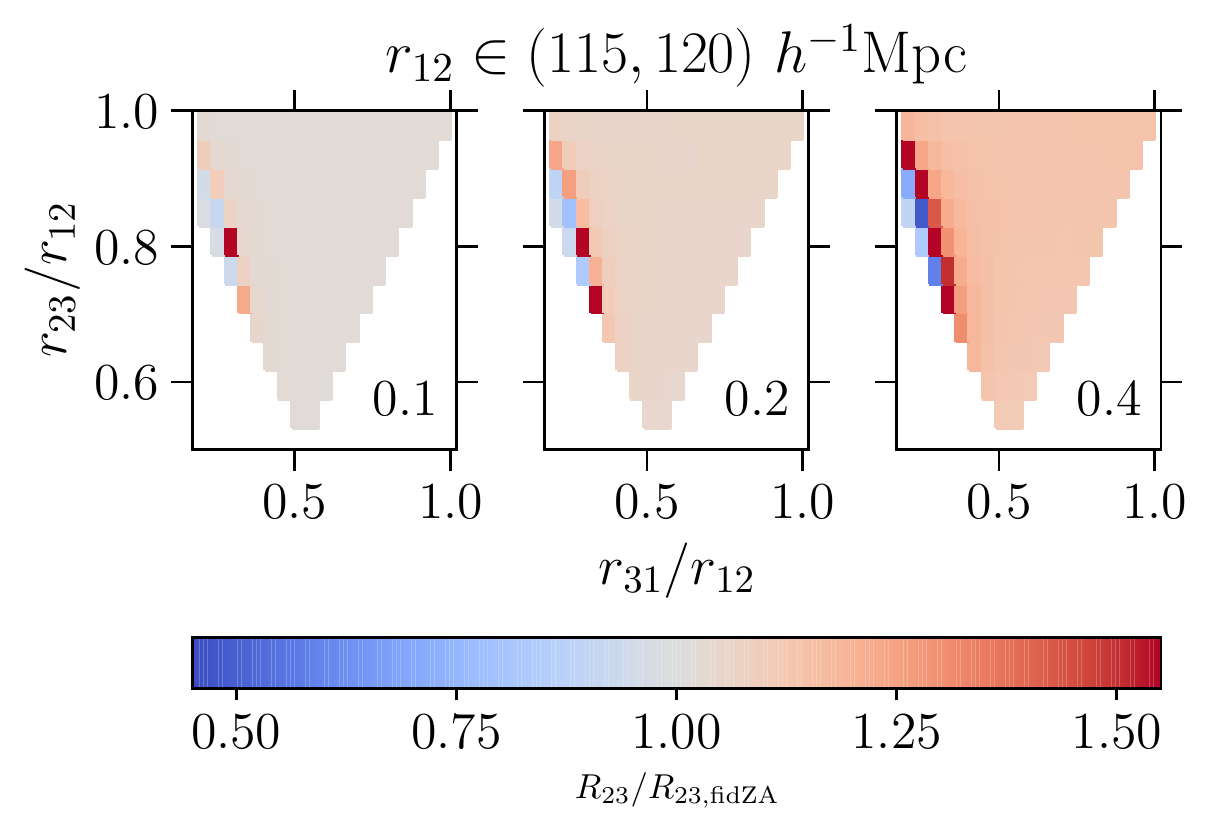}
  \caption{{\textbf{Top:} Effect of $M_\nu$ on $R^{\mathrm{h}}_{12}$. The length of the fixed leg of the triangle corresponds to $r_{12}\in (115,120)\ h^{-1}\mathrm{Mpc}$. Each panel corresponds to a different $M_\nu$, and it is indicated in the bottom right of each panel (in units of eV). The initial conditions of the fiducial cosmology simulations used in this figure were generated using the  Zel'dovich approximation. \textbf{Bottom:} Same as top row, but for $R^{\mathrm{h}}_{23}$.} }
  \label{fig:mnu_effect_12}
\end{figure}

\begin{figure*}
  \centering
  \includegraphics[scale=0.54]{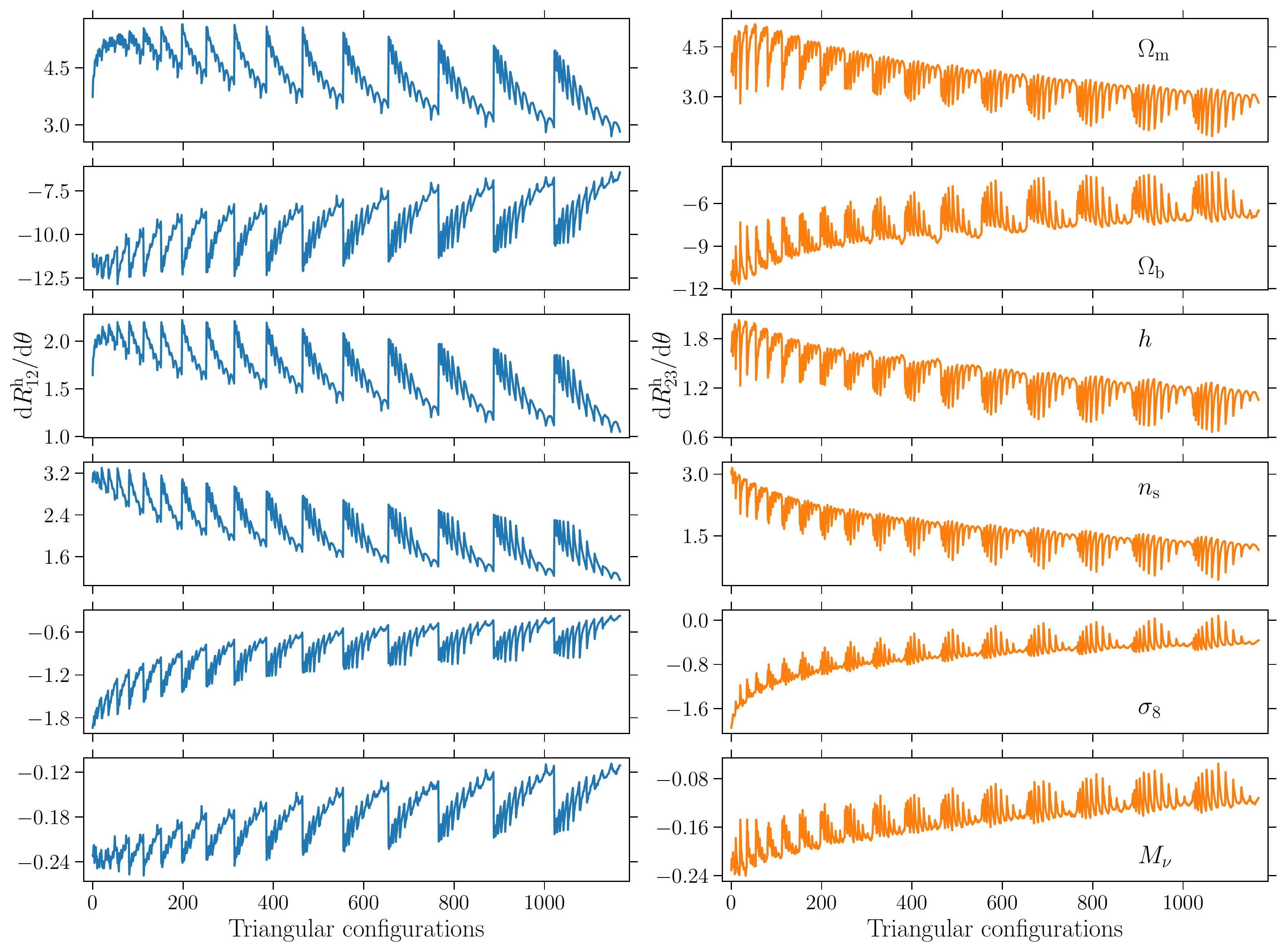} 
  \caption{Derivatives of $R^{\mathrm{h}}_{12}$ (left), and  $R^{\mathrm{h}}_{23}$ (right) with respect to the different cosmological parameters for all triangular configurations, in which the minimum and maximum separation for a triangular side is 20 and 120 $h^{-1}\mathrm{Mpc}$, respectively. The cosmological parameter under consideration is denoted on the right side of each row.} 
  \label{fig:derivatives}
\end{figure*}
In this section, we show how different cosmological parameters affect the relative velocities between halo pairs in a triplet. We focus on the shape dependence of $R^{\mathrm{h}}_{12}$ (top panel) and  $R^{\mathrm{h}}_{23}$ (bottom panel) on $M_\nu$ in Fig.~\ref{fig:mnu_effect_12}. The figure focuses on the triangular configurations in which the first leg of the triangle is fixed to $r_{12} \in (115, 120) \ h^{-1}\mathrm{Mpc}$.  In each panel, the top right part where $r_{31}/r_{12} = r_{23}/r_{12} = 1$ corresponds to equilateral configuration.  Bottom bins (i.e. $r_{31}/r_{12} = r_{23}/r_{12} \sim 0.5)$ contain the folded configurations. Since our analysis has a minimum separation length of $20 \ h^{-1}\mathrm{Mpc}$, we do not explore squeezed configurations ($r_{12} \sim r_{23} \gg r_{31}$). The left, middle, and right panels show the effect when the summed mass of neutrinos is 0.1, 0.2, and 0.4 eV, respectively. For the fiducial cosmology simulations, the 500 realisations of  which were generated using the Zel'dovich approximation (labelled `fidZA') were utilised. However, it will not make any difference in the ratio for these triangular configurations whether the simulations generated using second-order Lagrangian perturbation theory or Zel'dovich approximation are used, as we verified that both are numerically consistent with each other at $\sim 0.05\%$ at these scales. In the top panel, we see that by increasing the neutrino mass, the infall rate of $R^{\mathrm{h}}_{12}$ increases. The maximal difference as before is for the equilateral triangle, and minimal when $r_{31}\ll r_{23} < r_{12}$. In the case of $R^{\mathrm{h}}_{23}$, as seen from the bottom panel, the trend is similar when $r_{31}/r_{12} > 0.4$. However, for triangles that are close to squeezed configurations, the trend reverses, wherein on increasing the neutrino mass the infall rate of $R^{\mathrm{h}}_{23}$ decreases.

For the particular triangular configurations as shown in Fig.~\ref{fig:mnu_effect_12}, we  checked how other cosmological parameters affect the relative velocity statistics. We see that increasing $\Omega_{\mathrm{m}}$ decreases $R^{\mathrm{h}}_{12}$ and vice-versa. This overall behaviour is seen with variations in the Hubble parameter also, albeit with a different amplitude. However, for $\Omega_{\mathrm{b}}$ and $\sigma_8$, the behaviour is reversed, and the amplitude of the change of $R^{\mathrm{h}}_{12}$ with respect to its fiducial counterpart is within 6\% for these configurations. 
For the same triangular configurations, at $r_{31}/r_{12} \gtrsim 0.4$, the variation in cosmological parameters yields a similar response for $R^{\mathrm{h}}_{23}$ as $R^{\mathrm{h}}_{12}$, albeit with a different amplitude. However, this changes in the case of  $r_{31}/r_{12} < 0.4$, where in the behaviour reverses. 
The reverse is seen least in the case of variation in $\sigma_8$.
$R^{\mathrm{h}}_{12}$ and $R^{\mathrm{h}}_{23}$ show the maximal difference in response to cosmological parameters in the case of equilateral triangles and at $r_{31}/r_{12} < 0.4,$ respectively. We refer the reader to Appendix~\ref{app:shape} to see the figures showing how these different cosmological parameters affect the mean three-point relative velocity statistics.

When increasing the neutrino mass (as in the case of $M_\nu=0.4$ eV), there is a more pronounced shape dependence compared to $\sigma_8$. This should facilitate breaking the $M_\nu$-$\sigma_8$ degeneracy in two-point clustering information. 
To further understand this, we take a look at the derivatives for $R^{\mathrm{h}}_{12}$ (left panel) and  $R^{\mathrm{h}}_{23}$ (right panel) in Fig.~\ref{fig:derivatives} for all triangular configurations with  $r_{12} \geq r_{23} \geq r_{31} \geq 20\ h^{-1}\mathrm{Mpc}$ and a maximum separation of $120\ h^{-1}\mathrm{Mpc}$. Each row corresponds to the derivative with respect to a cosmological parameter (labelled in each row on the right panel). For the very large-scale triangles, the shape for $\partial R^{\mathrm{h}}_{23}/\partial \sigma_8$ and $\partial R^{\mathrm{h}}_{23}/\partial M_\nu$ looks similar. However, for the small-scale triangles, this is not the case. This points to the possibility that $R^{\mathrm{h}}_{23} + R^{\mathrm{h}}_{23}$ can be used to break the $M_{\nu}$-$\sigma_8$ degeneracy.

\subsection{Cosmological parameters}
\label{sec:cosmo_parameters}

\begin{figure*}
  \centering
  \includegraphics[scale=0.6]{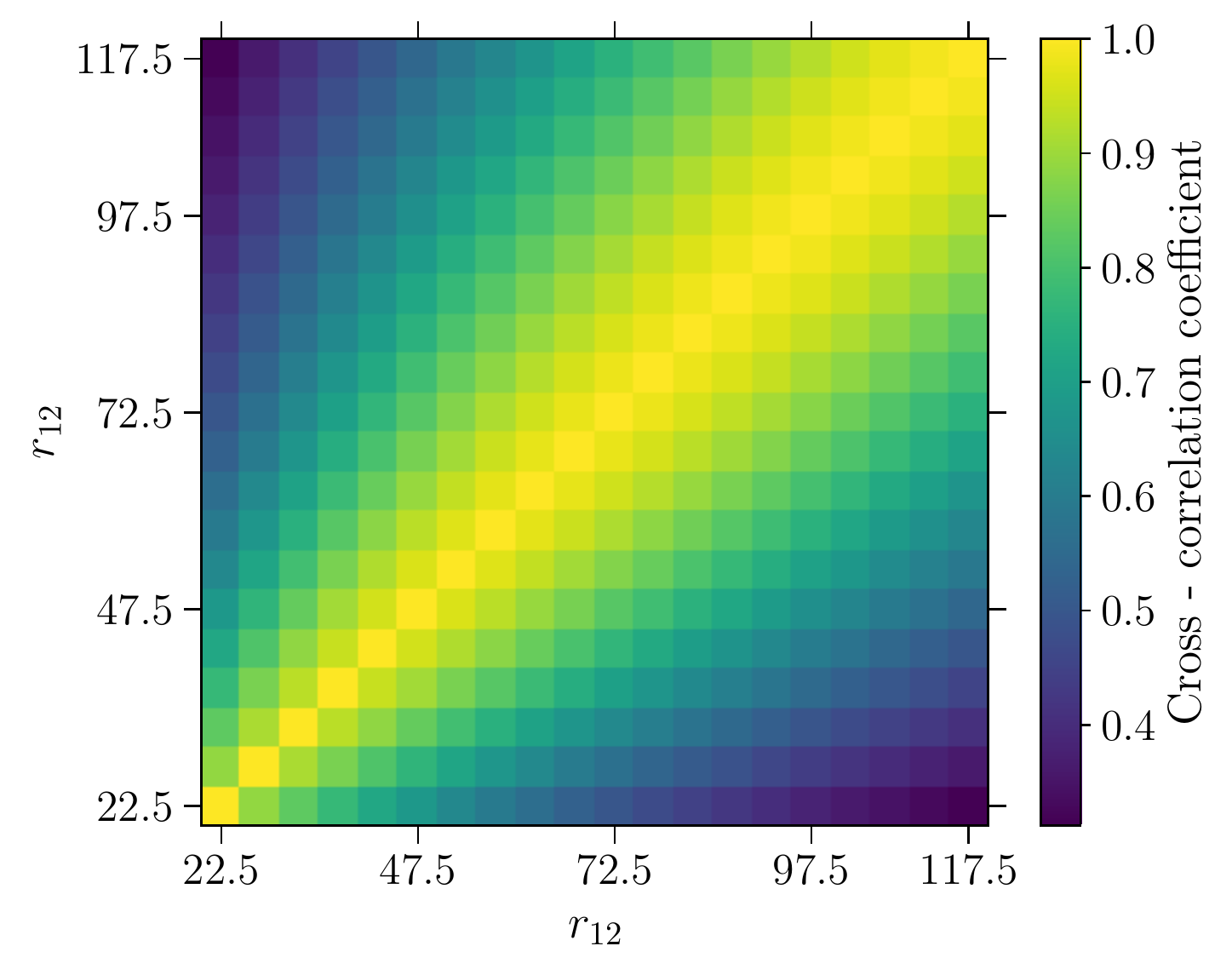} 
  \includegraphics[scale=0.6]{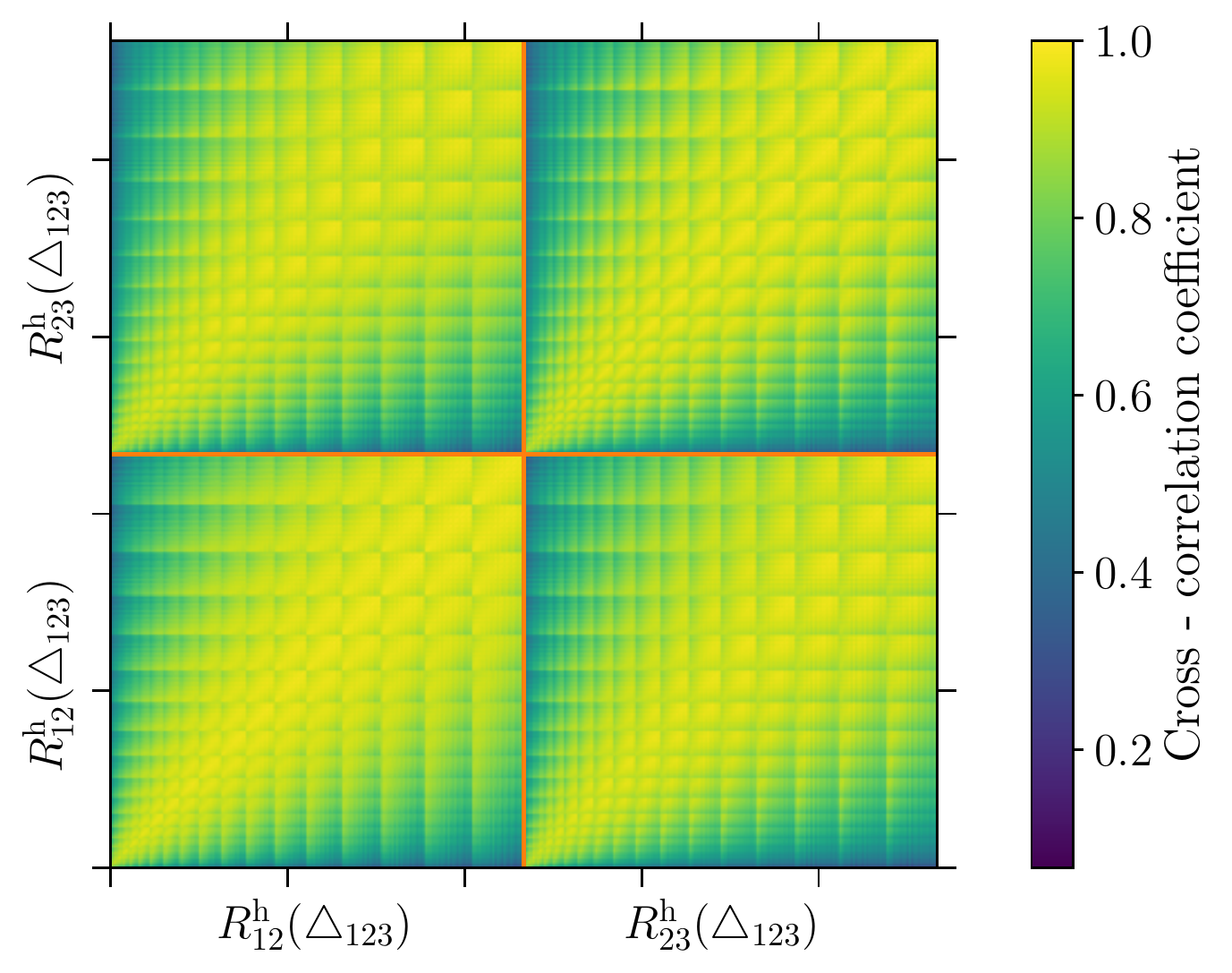} 
  \caption{Top panel shows correlation matrix in the case of halo pairwise velocities, while the bottom panel shows it for the relative velocities between halo pairs in a triplet computed using the 15000 realisations of the fiducial run of the Quijote suite of simulations. For the bottom panel, we used all triangular configurations with $20 \leq r_{31} \leq r_{23} \leq r_{12} \leq 120\ h^{-1}$Mpc.} 
  \label{fig:corr_three}
\end{figure*}

As can be seen from Eq.~\ref{eq:fisher_reduced}, one of the quantities entering the Fisher-matrix formalism is the inverse of the covariance matrix. It can be determined theoretically, in principle. For the mean pairwise velocities, for example, this was done in \cite{BhattacharyaKosowsky08} and \cite{Mueller+15a}. As the mean relative velocity between pairs in a triplet was introduced recently, no analytical prediction for its covariance matrix has been developed. We aim to tackle this in a future work. However in this work, we took the alternative route and measured the covariance matrix of the necessary summary statistics directly from the simulations using 15,000 realisations. Thus, in the case of $R^{\mathrm{h}}_{12}$ and  $R^{\mathrm{h}}_{23}$, this allowed us to compute the covariance matrix reliably. We show the cross-correlation coefficient ($\rho_{ij}=C_{ij}/\sqrt{C_{ii}C_{jj}}$) for both the mean halo pairwise velocity (left panel) and the mean relative velocity between halo pairs in a triplet (right panel) in Fig.~\ref{fig:corr_three}. We see that in both panels, it consists of prominent non-diagonal contributions. We especially notice that $R^{\mathrm{h}}_{12}$ and  $R^{\mathrm{h}}_{23}$ are highly correlated; this is not surprising, as in the leading order prediction we can see that they are a linear combination of different pieces of pairwise velocities of the three legs of the triangle (as  shown in Eqs.~\ref{eq:R12_triangle} and \ref{eq:R23_triangle}). 

The other quantities that enter Eq.~\ref{eq:fisher_reduced} are the derivatives of the model predictions with respect to the model parameters. The most common method to compute these derivatives is to use the theoretical predictions, such as those from the leading order in standard perturbation theory (Eq.~\ref{eq:bias}). We noted that these predictions are accurate at a few percent for triangular configurations above 50 $h^{-1}\mathrm{Mpc}$. At non-linear scales, the fidelity of the analytical predictions decreases. Hence, to circumvent this effect, we also used the derivatives measured directly from the simulations. We used 500 realisations, which were evaluated at thirteen different cosmologies, and each cosmological model with a small variation as already mentioned in Sect.~\ref{sec:sims}. 

\begin{figure*}
  \centering
  \includegraphics[scale=0.57]{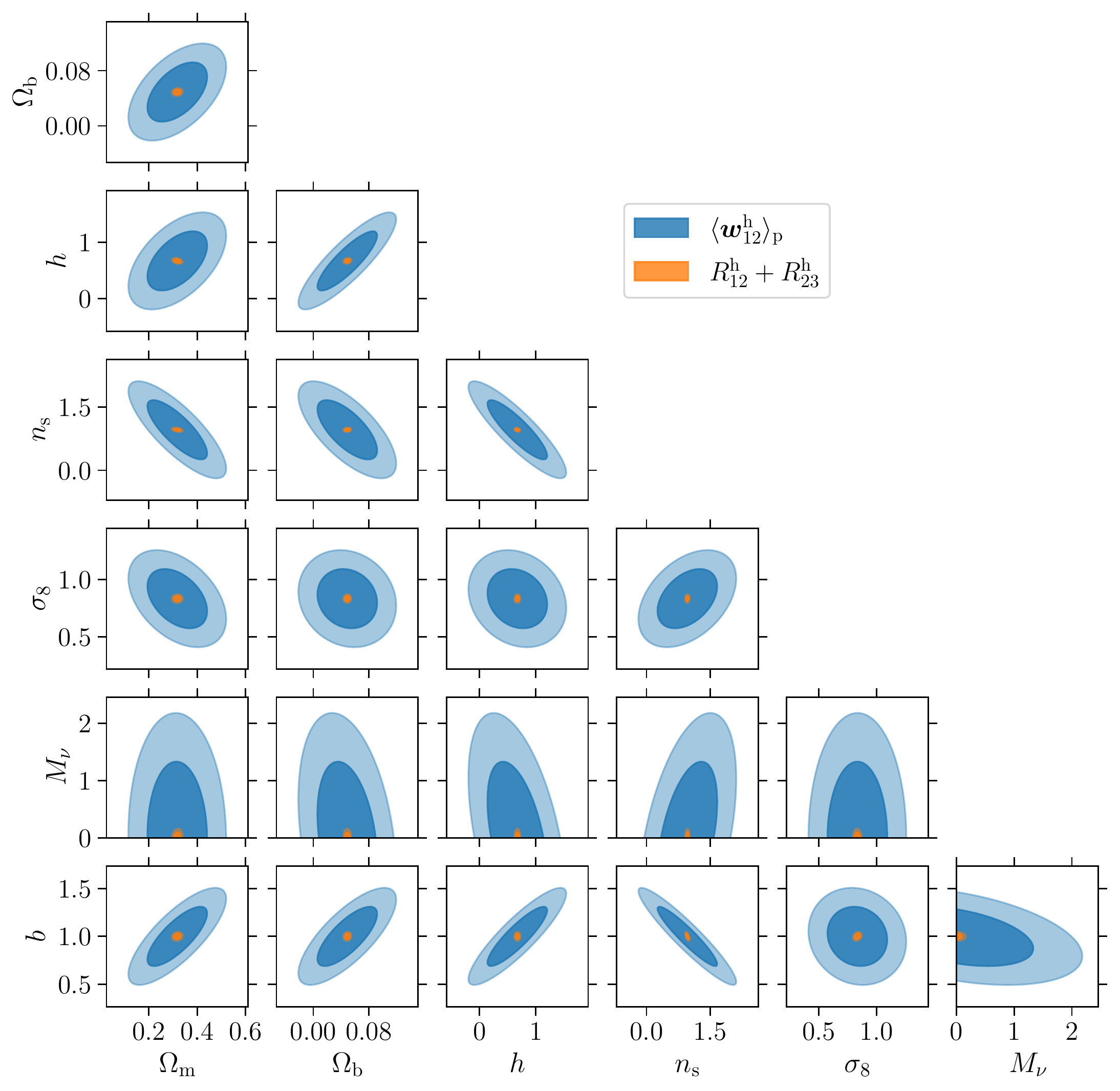}
  \caption{Joint 68.3\% (dark shaded contour) and 95.4\% (light shaded contour) credible region for all the pairs of model parameters (six cosmological and one nuisance parameter) at $z=0$. Colour-coding is indicated in the legend.} 
  \label{fig:fisher_results_combined}
\end{figure*}

\begin{figure}
  \centering
  \includegraphics[scale=0.41]{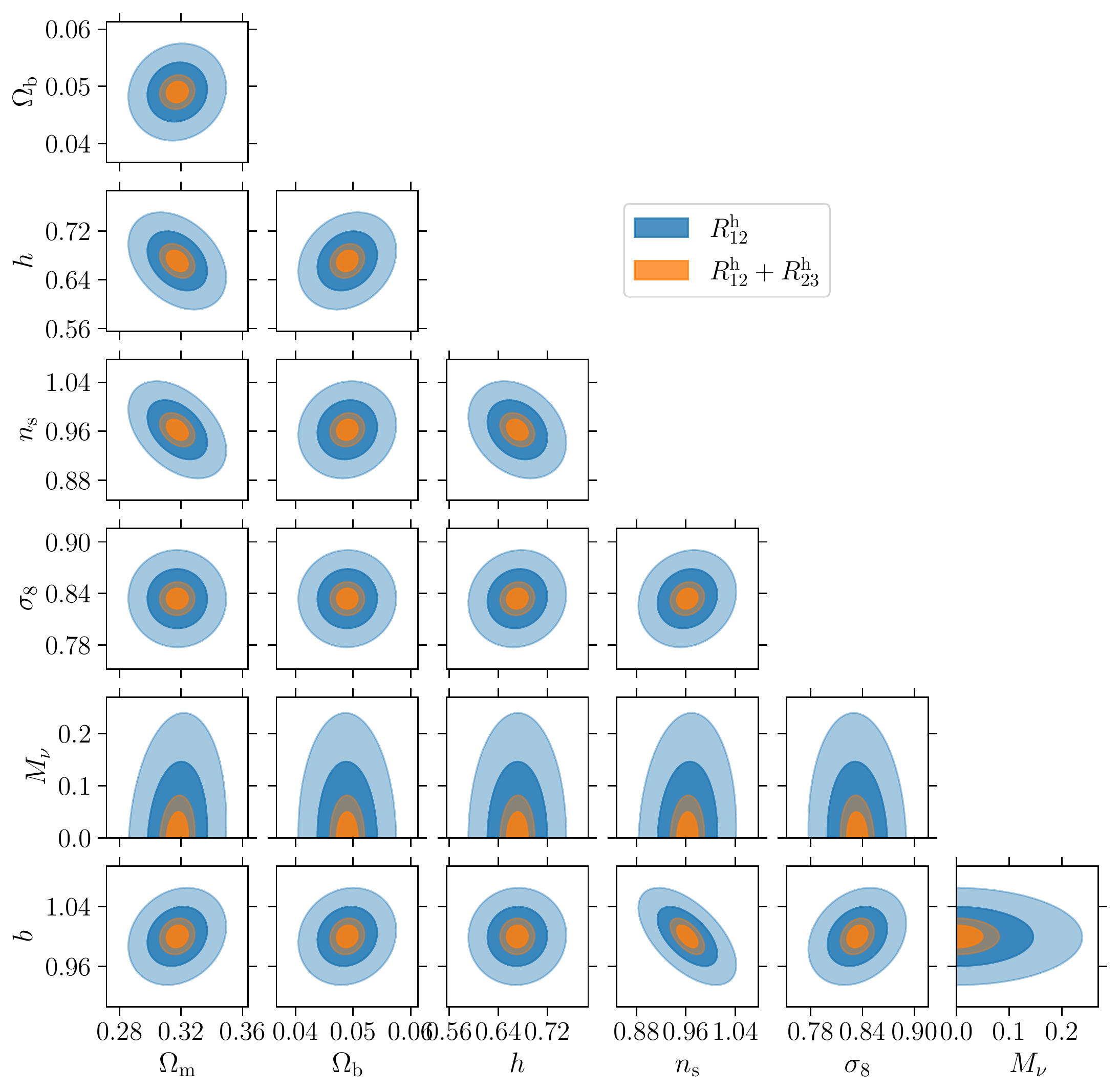} 
  \caption{Joint 68.3\% (dark shaded contour) and 95.4\% (light shaded contour) credible region for all the pairs of model parameters (six cosmological and one nuisance parameter) at $z=0$ for $R^{\mathrm{h}}_{12}$ (blue) and $R^{\mathrm{h}}_{12}+R^{\mathrm{h}}_{23}$ (orange), respectively.} 
  \label{fig:fisher_results_r12pluscombined}
\end{figure}

Combining both the necessary derivatives and the covariance matrix for the Fisher-matrix formalism, we present the constraints on the cosmological parameters in Fig.~\ref{fig:fisher_results_combined}. The 68.3 and 95.4 percent credible regions are given by dark and light shaded contour regions, respectively, for all possible pairs of model parameters obtained after marginalising over all the remaining parameters. The blue and orange ellipses show the constraints from the mean halo pairwise velocity and the mean relative velocity between halo pairs in a triplet, respectively. We can immediately note the benefits of constraining the cosmological parameters using the three-point relative velocities. We used the nuisance parameter $b$ as a scaling factor for $\langle \boldsymbol{w}^{\mathrm{h}}_{12} \rangle_{\mathrm{p}}$, $R^{\mathrm{h}}_{12}$, and  $R^{\mathrm{h}}_{23}$. For $r_{\mathrm{min}} = 20\ h^{-1}\mathrm{Mpc}$, we find that $R^{\mathrm{h}}_{12} + R^{\mathrm{h}}_{23}$ has a $1\sigma$ constraints for $\{\Omega_{\mathrm{m}}, \Omega_{\mathrm{b}}, h, n_{\mathrm{s}}, \sigma_8, M_\nu\}$ = $\{0.0091, 0.0024, 0.0226, 0.0221, 0.0156, 0.0655\ \mathrm{eV}\}$, which presents an improvement over constraints from $\langle \boldsymbol{w}^{\mathrm{h}}_{12} \rangle_{\mathrm{p}}$ by a factor of $\{8.91, 11.81, 15.46, 20.97, 10.93, 13.40\}$. The three-point mean halo relative velocity statistics thus offer a substantial improvement on the model parameter constraints over the mean halo pairwise velocity. To understand the role that the non-linear scales play in this improvement, we constrained the model parameters using Fisher matrix formalism where we restricted the triangular configurations at different $r_{\mathrm{min}}$ (keeping $r_{\mathrm{max}}$ fixed). We show the results in Table~\ref{tab:constraints}. For all the parameters considered, we see how  the constraints improve when going to smaller scales, as expected. The improvement obtained from $R^{\mathrm{h}}_{12} + R^{\mathrm{h}}_{23}$ over those obtained from mean pairwise velocity increases as the minimum separation scale considered decreases.

For the mean pairwise velocity, one can see that the spectral index $n_{\mathrm{s}}$ is highly degenerate with $\Omega_{\mathrm{m}}$,  $\Omega_{\mathrm{b}}$, and $h$. This degeneracy is also observed in the case of constraints from halo \citep{Hahn+20} and galaxy \citep{HahnVilla20} power spectrum. However, by considering the mean relative velocity between pairs in a triplet, the $n_{\mathrm{s}}$-$\Omega_{\mathrm{b}}$ degeneracy is broken. This is also supported by the fact that the impact of $n_{\mathrm{s}}$ on $R^{\mathrm{h}}_{12}$ and  $R^{\mathrm{h}}_{23}$ is different from those of $\Omega_{\mathrm{b}}$ as supported by Fig.~\ref{fig:derivatives} (also Figs.~\ref{fig:cosmo_effect_12} and \ref{fig:cosmo_effect_23}).

\begin{table*}
\caption{Marginalised 1$\sigma$ errors for the cosmological parameters and the nuisance parameter ($b$).}
\centering
\renewcommand{\arraystretch}{1.3}
\begin{tabular}{ccccccccccc}
\hline
 & \multicolumn{1}{c|}{} & \multicolumn{9}{c}{$r_{\mathrm{min}}\ (h^{-1}\mathrm{Mpc})$} \\ \cline{3-11} 
 & \multicolumn{1}{c|}{} & 20 & 25 & 30 & 35 & 40 & 45 & 50 & 55 & 60 \\ \hline \hline
\multirow{3}{*}{$\Omega_{\mathrm{m}}$} & $\langle \bm{w}_{12}|\bm{r}_{12} \rangle_\mathrm{p}$ & 0.08145 & 0.08375 & 0.08579 & 0.08580 & 0.09763 & 0.10089 & 0.10434 & 0.10440 & 0.11504 \\
 & $R^{\mathrm{h}}_{12} + R^{\mathrm{h}}_{23}$ & 0.00915 & 0.00963 & 0.01022 & 0.01087 & 0.01176 & 0.01277 & 0.01409 & 0.01568 & 0.01753 \\
 & F.I. & 8.90561 & 8.70010 & 8.39213 & 7.89261 & 8.30407 & 7.90352 & 7.40415 & 6.65691 & 6.56149 \\ \hline
\multirow{3}{*}{$\Omega_{\mathrm{b}}$} & $\langle \bm{w}_{12}|\bm{r}_{12} \rangle_\mathrm{p}$ & 0.02831 & 0.02997 & 0.03111 & 0.03119 & 0.03121 & 0.03132 & 0.03253 & 0.03259 & 0.03318 \\
 & $R^{\mathrm{h}}_{12} + R^{\mathrm{h}}_{23}$ & 0.00240 & 0.00249 & 0.00263 & 0.00278 & 0.00296 & 0.00323 & 0.00353 & 0.00393 & 0.00444 \\
 & F.I. & 11.81117 & 12.06080 & 11.81358 & 11.21327 & 10.53949 & 9.68702 & 9.20470 & 8.29658 & 7.47404 \\ \hline
\multirow{3}{*}{$h$} & $\langle \bm{w}_{12}|\bm{r}_{12} \rangle_\mathrm{p}$ & 0.34911 & 0.36354 & 0.38179 & 0.38180 & 0.38186 & 0.39210 & 0.42847 & 0.43300 & 0.43385 \\
 & $R^{\mathrm{h}}_{12} + R^{\mathrm{h}}_{23}$ & 0.02259 & 0.02377 & 0.02513 & 0.02648 & 0.02836 & 0.03079 & 0.03328 & 0.03644 & 0.04095 \\
 & F.I. & 15.45669 & 15.29114 & 15.19120 & 14.42058 & 13.46272 & 12.73427 & 12.87531 & 11.88329 & 10.59576 \\ \hline
\multirow{3}{*}{$n_{\mathrm{s}}$} & $\langle \bm{w}_{12}|\bm{r}_{12} \rangle_\mathrm{p}$ & 0.46385 & 0.47031 & 0.49449 & 0.49669 & 0.50729 & 0.55604 & 0.57261 & 0.57731 & 0.57766 \\
 & $R^{\mathrm{h}}_{12} + R^{\mathrm{h}}_{23}$ & 0.02212 & 0.02341 & 0.02492 & 0.02665 & 0.02875 & 0.03078 & 0.03397 & 0.03770 & 0.04278 \\
 & F.I. & 20.96929 & 20.08857 & 19.83974 & 18.63610 & 17.64347 & 18.06515 & 16.85804 & 15.31241 & 13.50433 \\ \hline
\multirow{3}{*}{$\sigma_8$} & $\langle \bm{w}_{12}|\bm{r}_{12} \rangle_\mathrm{p}$ & 0.17063 & 0.17985 & 0.18694 & 0.18736 & 0.26644 & 0.31528 & 0.36319 & 0.36372 & 0.37668 \\
 & $R^{\mathrm{h}}_{12} + R^{\mathrm{h}}_{23}$ & 0.01561 & 0.01657 & 0.01737 & 0.01841 & 0.01974 & 0.02176 & 0.02374 & 0.02625 & 0.02934 \\
 & F.I. & 10.93136 & 10.85548 & 10.76487 & 10.17951 & 13.49618 & 14.48853 & 15.29761 & 13.85747 & 12.83753 \\ \hline
\multirow{3}{*}{$M_\nu$} & $\langle \bm{w}_{12}|\bm{r}_{12} \rangle_\mathrm{p}$ & 0.87743 & 0.92138 & 0.97446 & 0.99740 & 1.04269 & 1.06858 & 1.22466 & 1.22523 & 1.33489 \\
 & $R^{\mathrm{h}}_{12} + R^{\mathrm{h}}_{23}$ & 0.06547 & 0.06828 & 0.07146 & 0.07585 & 0.08156 & 0.08845 & 0.09809 & 0.10836 & 0.12265 \\
 & F.I. & 13.40268 & 13.49481 & 13.63547 & 13.15010 & 12.78485 & 12.08153 & 12.48528 & 11.30666 & 10.88373 \\ \hline
\multirow{3}{*}{$b$} & $\langle \bm{w}_{12}|\bm{r}_{12} \rangle_\mathrm{p}$ & 0.20413 & 0.20429 & 0.20610 & 0.21088 & 0.22348 & 0.22854 & 0.30910 & 0.32395 & 0.34865 \\
 & $R^{\mathrm{h}}_{12} + R^{\mathrm{h}}_{23}$ & 0.01945 & 0.02333 & 0.02809 & 0.03340 & 0.04042 & 0.04755 & 0.05586 & 0.06514 & 0.07683 \\
 & F.I. & 10.49718 & 8.75640 & 7.33811 & 6.31302 & 5.52863 & 4.80646 & 5.53373 & 4.97336 & 4.53813 \\ \hline
\end{tabular}
\tablefoot{Different columns show the constraints obtained by changing the minimum separation length considered, while fixing the maximum separation to 120$\ h^{-1}\mathrm{Mpc}$. The various rows display the results for different parameters, and each of these rows is further divided into three rows to focus on the constraint coming from $\langle \bm{w}_{12}|\bm{r}_{12} \rangle_\mathrm{p}$, $R^{\mathrm{h}}_{12} + R^{\mathrm{h}}_{23}$, and the factor of improvement of the constraints from $R^{\mathrm{h}}_{12} + R^{\mathrm{h}}_{23}$ over $\langle \bm{w}_{12}|\bm{r}_{12} \rangle_\mathrm{p}$ denoted by  F.I., respectively.}
\label{tab:constraints}
\end{table*}

In this work, we made use of Fisher matrix formalism to study the informational content, where the derivatives and the covariance matrix was directly computed numerically from the Quijote suite of simulations. The numerical stability of our results stemming from the number of simulations used to compute the covariance matrix and derivatives is discussed in depth in Appendix~\ref{app:numericalstability}. We see that the constraints on the cosmological parameters vary less well below 1\% up on using 10,000 or more simulations for computing the covariance matrix. While in the case of the derivatives, the cosmological parameters are within around 5\% using 450 realisations when compared to the estimates using the full 500 realisations. The constraint on the nuisance parameter, $b$, varies less than 2.5\% and 5\% using 400 and 450 realisations, respectively. Hence, we conclude that our results are stable and devoid of any numerical inconsistencies.

So far, we considered the information content from combining $R^{\mathrm{h}}_{12}$ and $R^{\mathrm{h}}_{23}$,  but it is also worth exploring the constraining power of the two quantities when considered separately. We followed the same procedure as before to calculate the Fisher information matrix, and we computed the necessary derivatives and the covariance matrix directly from the simulations. We considered all triangular configurations with a minimum and a maximum separation of 20 $h^{-1}\mathrm{Mpc}$ and 120 $h^{-1}\mathrm{Mpc}$. In one case, we considered the constraints from $R^{\mathrm{h}}_{12}$ and $R^{\mathrm{h}}_{23}$ separately, and in the second case we consider the joint $R^{\mathrm{h}}_{12} + R^{\mathrm{h}}_{23}$ as has been done in the main text. We showcase the results in Fig.~\ref{fig:fisher_results_r12pluscombined}. The blue contours show the joint credible regions for all possible pairs of parameters using $R^{\mathrm{h}}_{12}$ alone, while the orange contour shows that from $R^{\mathrm{h}}_{12}+R^{\mathrm{h}}_{23}$. In the figure, we have not shown the joint credible regions from $R^{\mathrm{h}}_{23}$ because it is the same as $R^{\mathrm{h}}_{12}$. We can see that on all parameters $R^{\mathrm{h}}_{12}+R^{\mathrm{h}}_{23}$ is able to obtain tighter constraints. Quantitatively, they are 1.34 - 1.46 times better than those considering only $R^{\mathrm{h}}_{12}$ or $R^{\mathrm{h}}_{23}$.

In our main analysis, we did not incorporate any observation effects except for the bias term, and thus our results are provided with the ideal constraining power from these relative velocity statistics. In the case of kSZ experiments, one would have to additionally consider the optical depth as an additional free parameter, which depends on the halo mass \citep{Battaglia16,Flender+17}. We tried to (naively) mimic this halo mass dependence of optical depth by computing the derivative by changing the minimum halo mass considered in addition to the bias term and included it in the Fisher matrix analysis. We find that the constraining power on all six cosmological parameters remains unchanged at less than 0.1\%. However, the 1-$\sigma$ constraint on $b$ degrades from 0.019 to 0.022 (decrease by 12.7\%) after adding $M_{\mathrm{h},\mathrm{min}}$ as an additional parameter in the Fisher analysis. Accounting for additional observational effects related to the actual measurement of the kSZ signal such as beam convolution, contamination by other astrophysical signals, foreground cleaning, etc. would require an actual analysis based on kSZ-like mocks. This represents a quite different analysis that goes beyond the goals of the present work. 

We can compare our constraints with the ones obtained using multipoles of the power spectrum and bispectrum to see how well the three-point mean relative velocity statistics perform. The constraints from \cite{Hahn+20} provide us with an avenue to obtain a fair comparison of our summary statistics based on relative velocity with the clustering statistics, as both studies utilise the same halo catalogues. A caveat being the number density is slightly lower in our case as we chose a higher minimum mass of halos ($M_{\mathrm{h}} > 5\times10^{13}\ h^{-1}\mathrm{M}_\odot$, while in {\protect\NoHyper\citet{Hahn+20}\protect\endNoHyper} they considered $M_{\mathrm{h}} > 3.2\times10^{13}\ h^{-1}\mathrm{M}_\odot$). Using the monopole and quadrupole of the redshift-space power spectrum for $k_{\mathrm{max}}=0.5\ h\,\mathrm{Mpc}^{-1}$, {\protect\NoHyper\citet{Hahn+20}\protect\endNoHyper} were able to obtain 1-$\sigma$ constraint of 0.294 eV for the summed neutrino mass. Thus, our constraints on $M_\nu$ (i.e. 0.065 eV) from $R^{\mathrm{h}}_{12}+R^{\mathrm{h}}_{23}$ (for $r_{\mathrm{min}}=20\ h^{-1}\mathrm{Mpc}$) give an improvement factor of 4.5 over those obtained from the multipoles of the power spectrum. This improvement also crosses over to other cosmological parameters that both studies considered, with an improvement factor of (2.6, 4.8, 4.9, 5.7, 2.3) for $(\Omega_{\mathrm{m}}, \Omega_{\mathrm{b}}, h, n_{\mathrm{s}}, \sigma_8),$ respectively. 

{\protect\NoHyper\citet{Hahn+20}\protect\endNoHyper} also provided the cosmological constraints from using the redshift-space bispectrum monopole (the three-point clustering statistics in Fourier space). Our constraint on $M_\nu$ from the three-point relative velocity statistics is slightly less than the 1-$\sigma_8$ constraints from the bispectrum monopole (0.054 eV). However, for the other parameters such as $(\Omega_{\mathrm{m}}, \Omega_{\mathrm{b}}, h, n_{\mathrm{s}})$, we notice a small increase in the constraining power of three-point mean relative velocity when compared to those from bispectrum monopole with a factor of (1.2, 1.7, 1.7, 1.5), respectively. Albeit $\sigma_8$ produces a slight decrease in the constraining power with a factor of 0.9. This shows the potential of three-point mean relative velocity statistics to be competitive with respect to clustering information in the future.

%%%%%%%%%%%%%%%%%%%%%%%%
\section{Conclusion}
\label{sec:conclusions}
%%%%%%%%%%%%%%%%%%%%%%%%

\cite{KuruvillaPorciani20} introduced the mean relative velocity between pairs in a triplet, and derived the analytical prediction for it at leading order using standard perturbation theory. In this work, we quantified the information content in this three-point relative velocity statistics. Furthermore, we showed the fidelity of analytical the prediction for halos.
We accounted for the biasing by introducing a simple (scale-independent and mass-dependent) bias term. 
For triangular configurations with $r_{\mathrm{min}} = 50\ h^{-1}\mathrm{Mpc}$ and $r_{\mathrm{max}} = 120\ h^{-1}\mathrm{Mpc}$, we find that these predictions are accurate within a few percent. 

We checked the effect of cosmological parameters on the mean relative velocities between halo pairs in a triplet.
We find distinctive shape dependence on $R^{\mathrm{h}}_{12}$ and $R^{\mathrm{h}}_{23}$ as a result of varying cosmological parameters for the triangular configurations we considered. With regard to the summed mass of neutrinos, the shape dependence becomes more prominent as $M_\nu$ increases.

To quantify the information content in the two- and three-point mean relative velocity statistics, we used the Fisher-matrix formalism, where the necessary derivatives and the covariance matrices were directly measured from the Quijote suite of simulations. For the covariance matrices, we utilised 15,000 realisations of the reference cosmology.  The derivatives for the three-point mean relative velocity were also computed directly from the simulations. Combining these, we obtained the Fisher constraints by utilising the mean halo pairwise velocities and $R^{\mathrm{h}}_{12} + R^{\mathrm{h}}_{23}$. In the case of pairwise velocities, we used pair separations from 20 $h^{-1}\mathrm{Mpc}$ to 120 $h^{-1}\mathrm{Mpc}$, with a bin width of 5 $h^{-1}\mathrm{Mpc}$. In the case of $R^{\mathrm{h}}_{12}$ and $R^{\mathrm{h}}_{23}$, we considered all triangular configurations with a minimum triangle side length of 20 $h^{-1}\mathrm{Mpc}$ and a maximum of 120 $h^{-1}\mathrm{Mpc}$. This amounted to using 1168 total triangular configurations. $R^{\mathrm{h}}_{12} + R^{\mathrm{h}}_{23}$ was able to achieve a 1$\sigma$ constraint of 0.0655 eV for $M_\nu$, which is factor of 13.40 improvement over constraints from the mean pairwise velocities. The current lower limit of the sum of neutrino masses, $M_\nu = \sum m_\nu \gtrsim 0.06$ eV, comes from the neutrino oscillation experiments \citep[e.g.][]{Forero+14,Gonzalez-Garcia+16,Capozzi+17,Salas+17}. Thus, the three-point mean relative velocity statistics could help us to pinpoint the summed mass of neutrinos in the future. Similarly, for the other cosmological parameters, the three-point mean relative velocities made it possible to achieve similar information gains over pairwise velocity and can be seen quantitatively in Table~\ref{tab:constraints}. This information gain is possible as $R^{\mathrm{h}}_{12}$ and $R^{\mathrm{h}}_{23}$, being a three-point relative velocity statistics,  make more shape dependence variation possible, with different cosmological parameters compared to the mean pairwise velocity.  We did not study the effect of binning on the constraints from the mean relative velocity statistics, which uses a rather broad bin width and hence could lead to information loss, especially in the case of mean pairwise velocities. Though in this work our goal was to have a fair comparison of the information content of both the two-point and three-point mean relative velocities, and hence we kept the same bin width.

In summary, we quantitatively show that the mean radial relative velocity between pairs in a triplet as a cosmological observable at redshift zero could lead to a significant information gain in comparison to the mean radial pairwise velocity. The constraining power from $R^{\mathrm{h}}_{12} + R^{\mathrm{h}}_{23}$ is comparable to those obtained from the halo bispectrum measurement \citep{Hahn+20} as detailed in Sect.~\ref{sec:cosmo_parameters}. This represents the first step in understanding the viability of utilising the mean relative velocity between pairs in a triplet as a cosmological observable from either future peculiar velocity surveys or kinetic Sunyaev-Zel'dovich experiments.

%%%%%%%%%%%%%%%%%%%%%%%%%%%
\begin{acknowledgements}
%%%%%%%%%%%%%%%%%%%%%%%%%%%
We thank the referee for their constructive comments, which helped in improving the manuscript. JK likes to thank Francisco Villaescusa-Navarro for all the valuable and helpful discussions regarding the Quijote simulation suite. We thank Julien Grain, Tony Bonnaire and Hideki Tanimura for carefully reading the manuscript, and providing comments. We acknowledge funding for the ByoPiC project from the European Research Council (ERC) under the European Union's Horizon 2020 research and innovation program grant agreement ERC-2015-AdG 695561. 
We are thankful  to  the  community for developing  and  maintaining open-source software packages extensively used in our work, namely \textsc{Cython} \citep{cython}, \textsc{Matplotlib} \citep{matplotlib} and \textsc{Numpy} \citep{numpy}.
\end{acknowledgements}

%%%%%%%%%%%%%%%%%%%%%%%%%%%%%%
\setlength{\bibhang}{2.0em}
\setlength\labelwidth{0.0em}
\bibliographystyle{aa}
\bibliography{main}
%%%%%%%%%%%%%%%%%%%%%%%%%%%%%%%

\begin{appendix}

\section{Numerical stability}
\label{app:numericalstability}

The Fisher matrix (Eq.~\ref{eq:fisher_reduced}) calculation depends on two elements: (i) the covariance matrix, and (ii) the derivatives of the summary statistics with respect to the cosmological parameters (and nuisance parameters).  In this work, we calculated both directly using the Quijote suite of simulations. Hence it is of vital importance to verify that we are not biased by any numerical effects. We explain how we tested the convergence of both these ingredients in this section.
\begin{figure}
  \centering
  \includegraphics[scale=0.58]{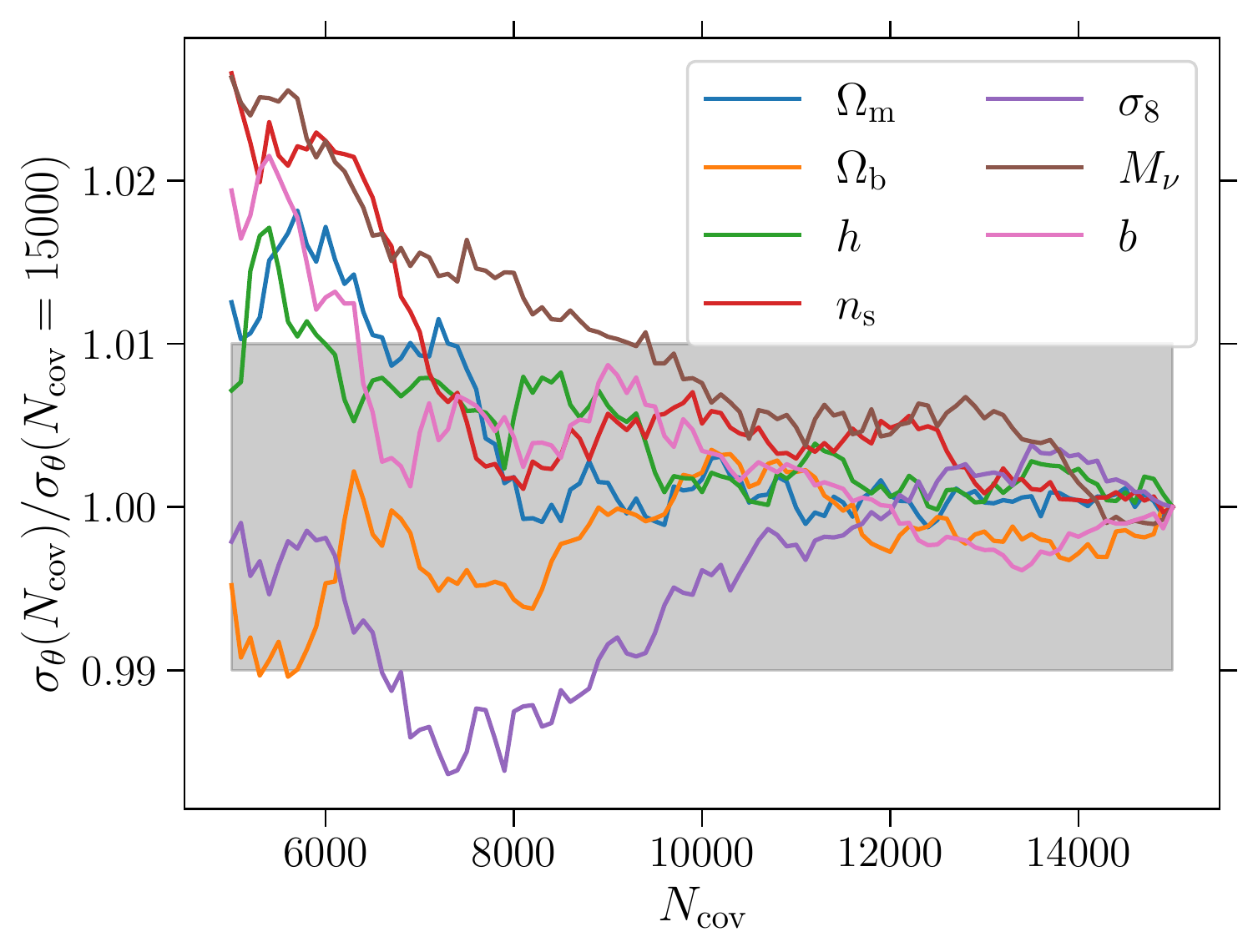}
  \caption{Convergence test of the marginalised one-sigma constraints by varying the number of simulations to compute the covariance matrix. The parameters in consideration are shown in the legend. The grey horizontal band shows the region where the variation in the constraints are within 1\% compared to the constraint obtained using $N_{\mathrm{cov}}=15000$.} 
  \label{fig:convergence}
\end{figure}

\begin{figure}
    \centering
  \includegraphics[scale=0.58]{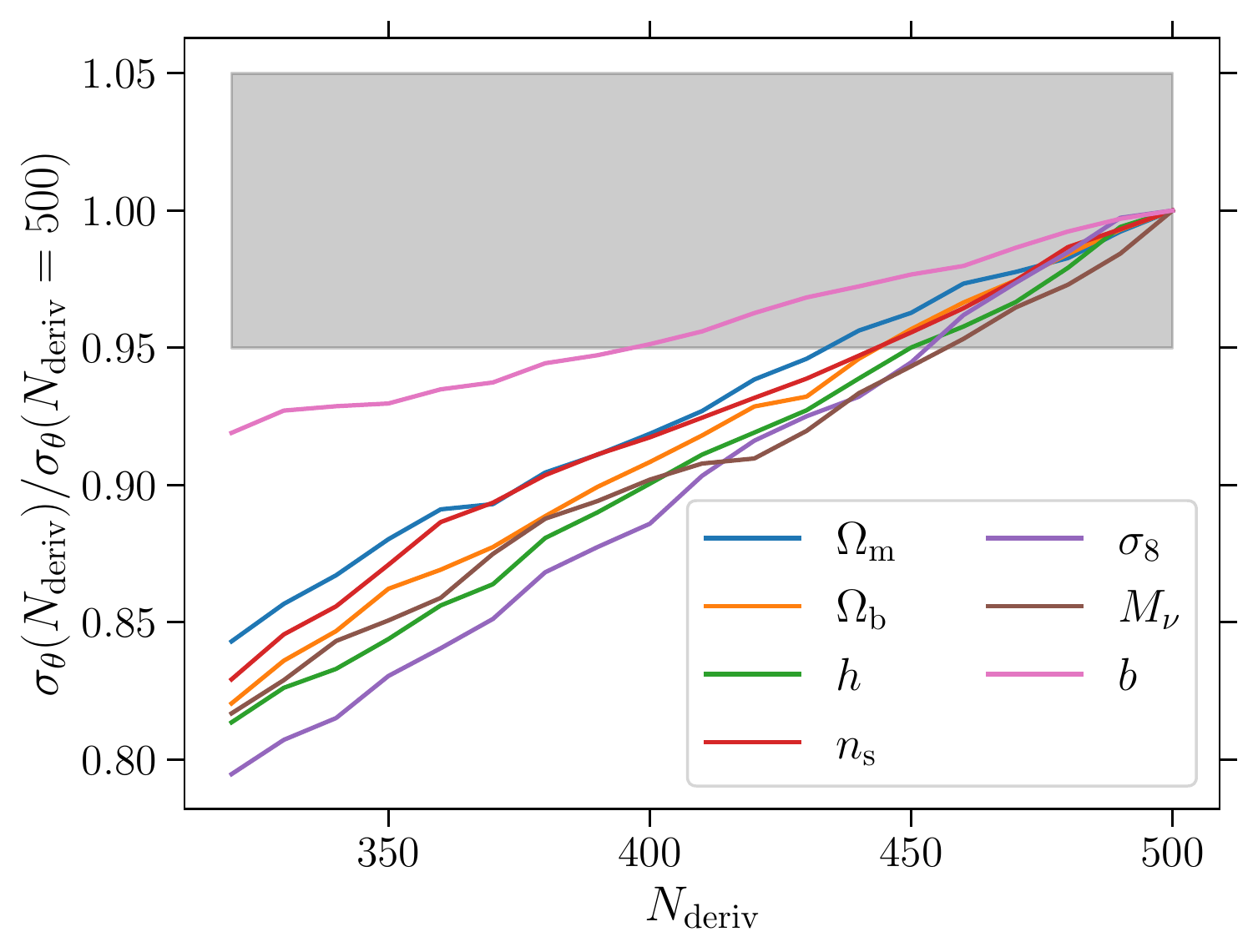} \\
  \includegraphics[scale=0.58]{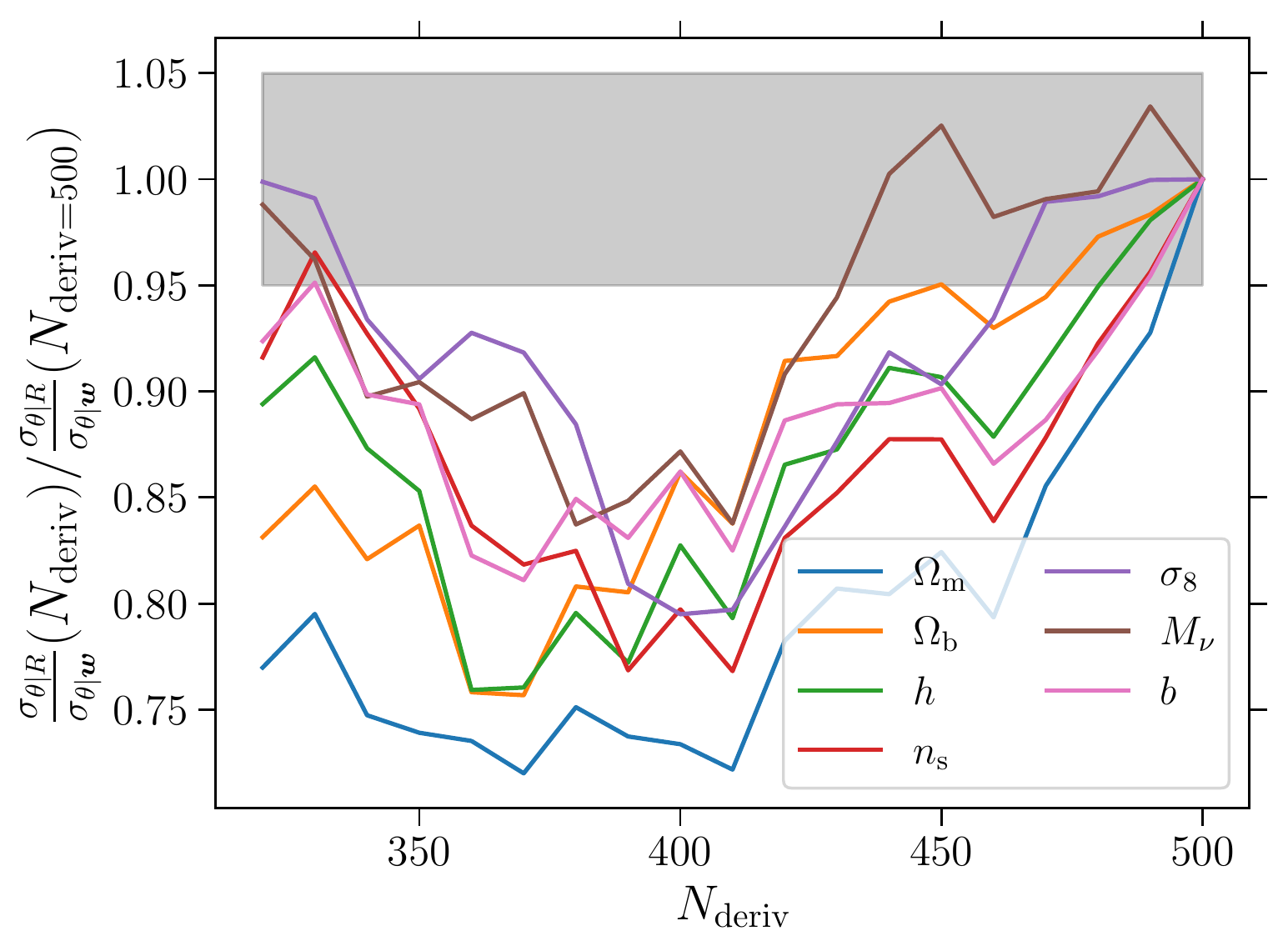}
    \caption{Marginalised one-sigma constraints by varying the number of simulations to compute the derivatives (top panel). Bottom panel shows the relative improvement of $R^{\mathrm{h}}_{12}+R^{\mathrm{h}}_{23}$ over $\langle \bm{w}_{12}|\bm{r}_{12} \rangle_\mathrm{p}$. The parameters in consideration are the six cosmological parameters and one nuisance parameter ($b$) as shown in the legend. The grey horizontal band denotes the region where the variation in the constraints are within 5\% compared to the constraint obtained using $N_{\mathrm{deriv}} = 500$.}
    \label{fig:num_derivatives}
\end{figure}

The covariance matrix was calculated using 15,000 realisations from the Quijote suite of simulations. However, the data vector of $R^{\mathrm{h}}_{12} + R^{\mathrm{h}}_{23}$ has $2\times1168$ triangular configurations. To verify that our analysis was not affected by numerical effects while calculating the covariance matrix, we varied the number of simulations and checked whether this impacted the final constraints from the Fisher forecast. In Fig.~\ref{fig:convergence}, we showcase the ratio between the marginalised 1$\sigma$ constraints (for all the cosmological parameters and the nuisance parameter considered in this work) obtained using covariance matrix calculated from $N_{\mathrm{cov}}$ simulations, and that obtained using covariance matrix calculated from 15,000 simulations. Using 5000 simulations to compute the covariance matrix leads to \~2-3\% on $n_{\mathrm{s}}$ and $M_{\nu}$ and roughly 1\% for $\Omega_{\mathrm{m}}, \Omega_{b}$ and $h$. However, when using 10,000 simulations or more, it can be seen that there is a clear convergence and a variation well below one percent on all parameter constraints (including the nuisance parameter). 

On the top panel of Fig.~\ref{fig:num_derivatives}, we show the ratio of the parameter constraint obtained using $N_{\mathrm{deriv}}$ realisations for derivatives to that obtained using $N_{\mathrm{deriv}} = 500$ realisations. We see that only using 300 realisations results in about a 20\% difference when compared to the constraints obtained using full 500 simulations. The variations are around 10\% and 5\% when using 400 and 450 realisations respectively. However, the nuisance parameter $b,$ which is a scaling factor, has the least variation when changing the number of simulations used for computing the derivatives. The variation is only around 2.5\% when using 450 simulations. In the bottom panel we show the relative improvement for the cosmological parameters form using $R^{\mathrm{h}}_{12} + R^{\mathrm{h}}_{23}$ over $\langle \bm{w}_{12}|\bm{r}_{12} \rangle_\mathrm{p}$. Using over 450 realisations of the simulations, the improvements remain mostly within 10\%.

\section{Shape dependence of   \texorpdfstring{$R^{\mathrm{h}}_{12}$}{R12} and  \texorpdfstring{$R^{\mathrm{h}}_{23}$}{R23}}
\label{app:shape}

\begin{figure}
  \centering
  \includegraphics[scale=0.56]{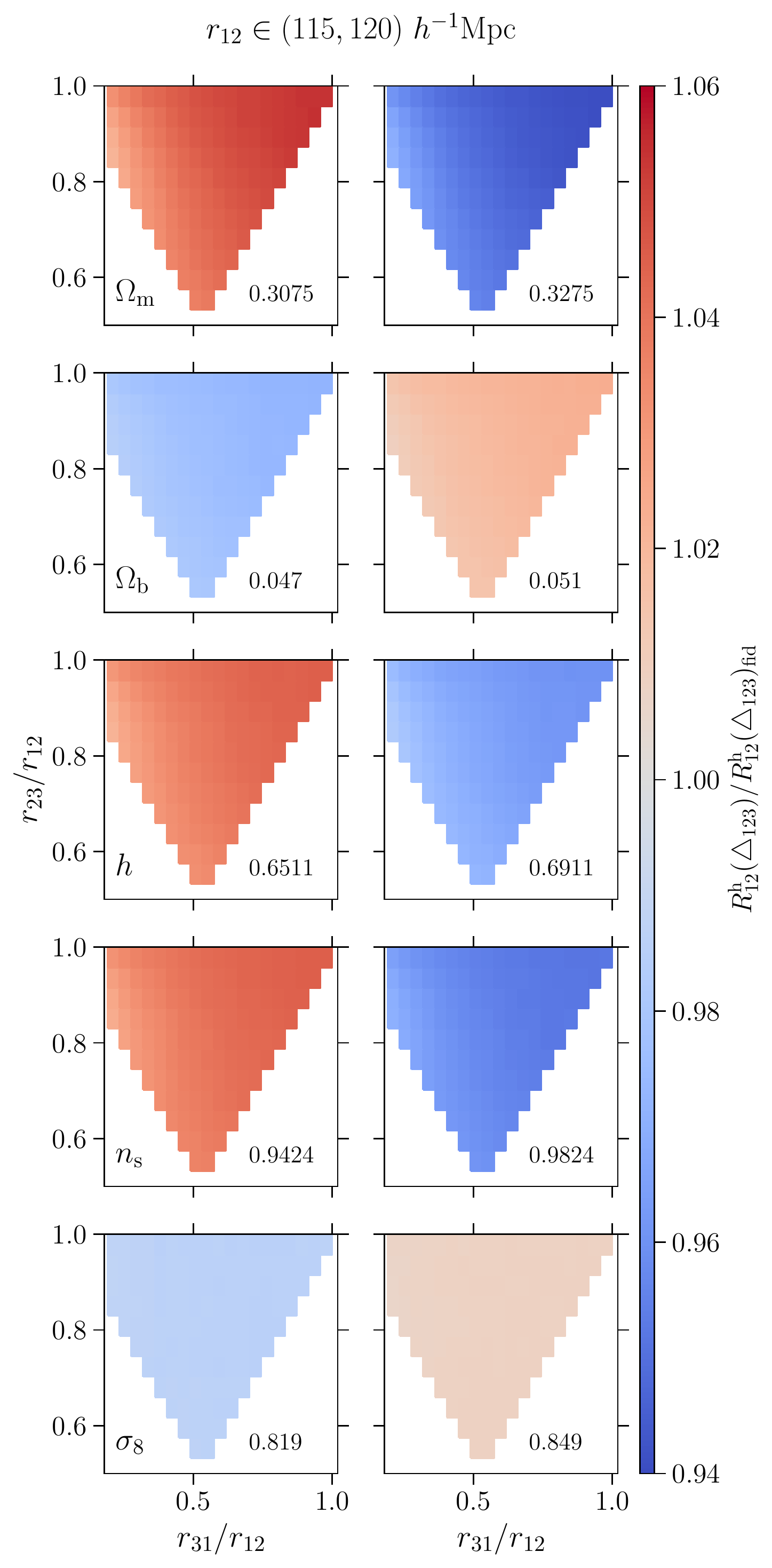} 
  \caption{Ratio between $R^{\mathrm{h}}_{12}(\triangle_{123})$ for varying cosmology and the fiducial cosmology. Each row corresponds to a different cosmological parameter, as given in the bottom left of the first column in each row. The length of the fixed leg of the triangle corresponds to $r_{12}\in (115,120)\ h^{-1}\mathrm{Mpc}$.} 
  \label{fig:cosmo_effect_12}
\end{figure}

\begin{figure}
  \centering
  \includegraphics[scale=0.56]{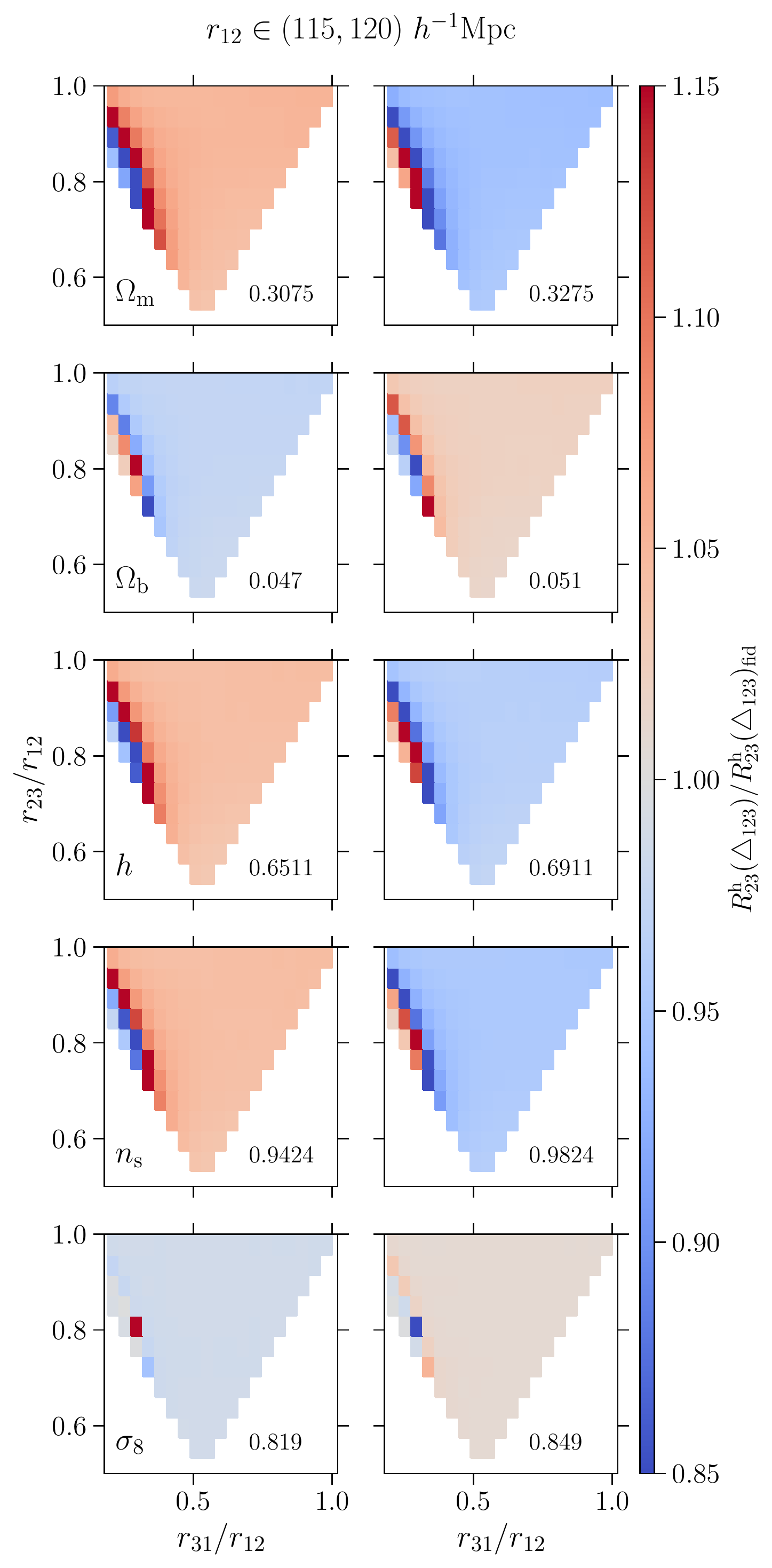} 
  \caption{Same as Fig.~\ref{fig:cosmo_effect_12}, but for $R^{\mathrm{h}}_{23}(\triangle_{123})$.} 
  \label{fig:cosmo_effect_23}
\end{figure}

In this section, we show how the cosmological parameters affect the mean three-point relative velocity statistics for the triangular configuration where one side of the leg, i.e. $r_{12}$ is fixed to have a pair separation between 115 and 120 $h^{-1}\mathrm{Mpc}$. Figs.~\ref{fig:cosmo_effect_12} and \ref{fig:cosmo_effect_23} show the shape dependence of $R^{\mathrm{h}}_{12}$ and $R^{\mathrm{h}}_{23}$ , respectively. Each row corresponds to variation of a single cosmological parameter. 

\end{appendix}
\end{document}